\documentclass[prd,10pt,preprintnumbers,
               twocolumn,numbers,sort&compress,nofootinbib,
               floatfix]{revtex4}
\usepackage{amssymb}  % \Box etc: see /opt/texmf/tex/ams/doc/amsguide.ps
\usepackage{amsfonts} % \mathfrak and \mathbb
\usepackage{amsmath}

\newcommand{\abs}[1]{\lvert{#1}\rvert}
\newcommand{\braket}[1]{\mathinner{\langle{#1}\rangle}}{\catcode`\|=\active
  \gdef\Braket#1{\left<\mathcode`\|"8000\let|\bravert {#1}\right>}}
\renewcommand{\d}[1][{\negthickspace}]{\mathrm{d}{#1}\;}

\newcommand{\ket}[1]{\mathinner{|{#1}\rangle}}
\newcommand{\mat}[1]{\boldsymbol{\mathbf{#1}}}
\newcommand{\norm}[1]{\lVert{#1}\rVert}
\newcommand{\pdiff}[3][{}]{\frac{\partial^{#1}{#2}}{\partial{#3}^{#1}}}
\newcommand{\vect}[1]{\vec{\mathbf{#1}}}

\DeclareMathOperator{\tr}{Tr}

\DeclareMathOperator{\order}{O}

\usepackage{psfrag}
\usepackage{graphicx}

\begin{document}
\date{30 November 2005}
\title{Kaon Condensation in a Nambu--Jona-Lasinio (NJL) Model at High Density}
\author{Michael McNeil Forbes}
\affiliation{Center for Theoretical Physics, Department of Physics,
  MIT, Cambridge, Massachusetts 02139}

\preprint{MIT-CTP-3554}
\begin{abstract}
  We demonstrate a fully self-consistent microscopic realization of a
  kaon-condensed colour-flavour locked state (CFLK$^0$) within the
  context of a mean-field NJL model at high density.  The properties
  of this state are shown to be consistent with the QCD low-energy
  effective theory once the proper gauge neutrality conditions are
  satisfied, and a simple matching procedure is used to compute the
  pion decay constant, which agrees with the perturbative QCD result.
  The NJL model is used to compare the energies of the CFLK$^0$ state
  to the parity even CFL state, and to determine locations of the
  metal/insulator transition to a phase with gapless fermionic
  excitations in the presence of a non-zero hypercharge chemical
  potential and a non-zero strange quark mass.  The transition points
  are compared with results derived previously via effective theories
  and with partially self-consistent NJL calculations.  We find that
  the qualitative physics does not change, but that the transitions
  are slightly lower.
\end{abstract}
\maketitle
\section{Introduction}
Recently there has been interest in the structure of matter at
extremely high densities, such as might be found in the cores of
neutron stars.  At large enough densities, the nucleons are crushed
together and the quarks become the relevant degrees of freedom.  The
asymptotic freedom of QCD ensures that the theory is weakly coupled at
high enough densities.  This allows one to perform weak-coupling
calculations at asymptotically high densities.  Such calculations have
established that the structure of the ground state of quark matter is
a colour superconductor (see for
example~\cite{Barrois:1977xd,Frautschi:1978,Barrois:1979,Bailin:1984bm,
  Alford:1998zt,Rapp:1998zu,Alford:1998mk,Son:1998uk,hong:1999ru,Evans:1999at,Rajagopal:2000wf,Alford:2001dt,Nardulli:2002ma,Reddy:2002ri,Schafer:2003vz}).
In particular, at densities high enough that the three lightest quarks
can be treated as massless, the ground state is the
colour-flavour--locked (CFL) state in which all three colours and all
three flavours participate in maximally (anti)-symmetric
pairing~\cite{Alford:1998mk,Schafer:1999fe,Evans:1999at,Shovkovy:1999mr}.

Determination of the QCD phase structure at moderate densities and in
the presence of non-zero quark masses has proceeded in several ways.
One approach has been to formulate a chain of effective theories, and
then to match coefficients across several energy scales through these
effective theories to perturbative calculations.  Coefficients in the
low-energy chiral effective theory~\cite{Casalbuoni:1999wu} are
matched to calculations performed in high-density effective theories
(HDET)~\cite{hong:1998tn,hong:1999ru,Beane:2000ms} which in turn are
matched to weakly-coupled QCD.  This allows one to determine the
properties of the Goldstone bosons and determine the effects of small
quark
masses~\cite{Miransky:1999tr,Son:1999cm,Son:2000tu,Rho:1999xf,hong:1999ei,Manuel:2000wm,Rho:2000ww,Zarembo:2000pj,hong:2000ng,Schafer:2001za}.
Within this framework, it has been noted that, in the presence of a
finite strange quark mass, neutral ``kaons'' (the lightest
pseudo-Goldstone modes at high density with the same quantum numbers
as their vacuum counterpart) can Bose-condense in the CFL state to
form a kaon-condensed CFLK$^0$ phase with lower condensation
energy~\cite{Schafer:2000ew,Bedaque:2001je,Kaplan:2001qk}.

Unfortunately, the low-energy effective theory is only reliable for
small perturbations and at moderate densities the strange quark mass
is not a small perturbation.  A recent attempt has been made to
extrapolate to large strange quark mass
($m_s$)~\cite{Kryjevski:2004kt}, but this approach has not dealt with
additional complications in the condensate structure that allow
different gap parameters for each pair of quarks.

To deal with moderate quark masses, another approach has been to study
Nambu--Jona-Lasinio (NJL) models~\cite{Nambu:1961tp,Nambu:1961fr} of
free quarks with contact interactions that model instanton interactions
or single gluon exchange.  These models are amenable to a mean-field
treatment and exhibit a similar symmetry breaking pattern to QCD which
results in CFL ground states~\cite{Alford:1998zt,Rapp:1998zu}.

Within these models, one can study the effects of moderate quark
masses through self-consistent solutions of the mean-field gap
equations.  This has led to a plethora of phases.  In particular,
several analyses show a transition to a colour-flavour locked phase
with gapless fermionic excitations (the gCFL phase).  These include
both NJL-based
calculations~\cite{Alford:2003fq,Alford:2004hz,Ruster:2004eg,Fukushima:2004zq}
and effective-theory--based
calculations~\cite{Kryjevski:2004jw,Kryjevski:2004kt}.  Until
recently, however, the NJL calculations have excluded the possibility
of kaon condensation (see however~\cite{Barducci:2004nc} which
considers kaon condensation in the NJL model at low density), while
the effective theories do not consider the complicated patterns in
which the condensate parameters evolve at finite quark masses.

The goal of this paper is to show that one can combine the analysis of
the low-energy effective theories, which exhibit kaon condensation,
with the self-consistent mean-field analysis of the NJL model, which
accounts for the full condensate structure.  In particular, we use an
NJL model based on single gluon exchange to find self-consistent
solutions that correspond to the CFLK$^0$ phase; we show that these
phases agree with the predictions of the low-energy effective theory;
and we determine how and where the zero temperature phase transition
to a gapless CFL phase occurs as one increases the strange quark mass.
In addition, unlike previous work on the NJL model, our numerical
solutions are \emph{fully} self-consistent: we include \emph{all}
condensates and self-energy corrections required to close the gap
equations.

We first describe the pattern of symmetry breaking that leads to the
CFL and CFLK$^0$ states (Section~\ref{sec:colo-flav-lock}).  Then we
present our numerical results, demonstrating some properties of these
states and determining the locations of the zero-temperature phase
transitions (Section~\ref{sec:self-cons-solut}).  After a careful
description of our model (Section~\ref{sec:njl-model}) we derive the
low-energy effective theory, paying particular attention to the
differences between QCD and the NJL model
(Section~\ref{sec:low-energy-effective}).  Here we demonstrate that,
for small perturbations, our numerical solutions are well described by
the effective theory, and we use our numerical results to compute the
pion decay constant $f_\pi$ which agrees with the perturbative QCD
results.  Specific numerical details about our calculations and a full
description of our self-consistent parametrization are given in
Appendix~\ref{sec:full-param}.

We leave for future work the consideration of finite temperature
effects, the analysis of the gapless CFLK$^0$ (gCFLK$^0$), the
inclusion of instanton effects, the inclusion of up and down quark
mass effects, and the possibility of other forms of meson condensation.

\section{Colour Flavour Locking (CFL)}
\label{sec:colo-flav-lock}
QCD has a continuous symmetry group of U$(1)_{\text{B}}\otimes
\text{SU}(3)_{\text{L}}\otimes \text{SU}(3)_{\text{R}}\otimes
\text{SU}(3)_{\text{C}}$.  In addition, there is an approximate
U$(1)_{\text{A}}$ axial flavour symmetry that is explicitly broken by
anomalies.  At sufficiently high densities, however, the instanton
density is suppressed and this symmetry is approximately restored.

The CFL ground state spontaneously breaks these continuous symmetries
through the formation of a diquark condensate~\cite{Alford:1998mk}
\begin{equation}
  \label{eq:diquark}
  \braket{\overline{\psi}^{C}_{\alpha a}\mat{\gamma}_5 \psi_{\beta b}} \propto 
  \Delta_3\epsilon^{\alpha \beta k}\epsilon_{abk}+
  \Delta_6(\delta^{\alpha}_{a}\delta^{\beta}_{b}+
  \delta^{\alpha}_{b}\delta^{\beta}_{a}).
\end{equation}
The symmetry breaking pattern (including the restored axial
U$(1)_{\text{A}}$ symmetry) is thus\footnote{The $\text{Z}_{3}$ factor
  mods out the common centres.  See~(\ref{eq:ExplicitSym}) for the
  explicit representation.}
\begin{equation}
  \label{eq:SSB}
  \frac{\text{U}(3)_{\text{L}} 
    \otimes \text{U}(3)_{\text{R}}
    \otimes \text{SU}(3)_{\text{C}}}
  {\text{Z}_{3}}
  \rightarrow 
  \text{SU}(3)_{\text{L}+\text{R}+\text{C}}\otimes\text{Z}_2\otimes\text{Z}_2
\end{equation}
where the $\text{Z}_2$ symmetries correspond to $\psi_L \rightarrow
-\psi_L$ and $\psi_R\rightarrow -\psi_R$.  It has been noted that the
symmetry breaking pattern at high density~(\ref{eq:SSB}) is the same
as that that for hyper-nuclear matter at low
density~\cite{Schafer:1998ef}.  This leads one to identify the
low-energy pseudo-scalar degrees of freedom in both theories.  We
shall refer to the pseudo-scalar Goldstone bosons in the high-density
phase as ``pions'' and ``kaons'' etc. when they have the same flavour
quantum numbers as the corresponding low-density particles.

The CFL state~(\ref{eq:diquark}) preserves parity, and is preferred
when instanton effects are considered.  Excluding instanton effects,
there is an uncountable degeneracy of physically equivalent CFL ground
states that violate parity.  These are generated from the parity even
CFL by the broken symmetry generators.

The symmetry breaking pattern~(\ref{eq:SSB}) breaks $18$ generators.
The quarks, however, are coupled to the eight gluons associated with
the SU$(3)_{\text{C}}$ colour symmetry and to the photon of the
U$(1)_{\text{EM}}$ electromagnetism (which is a subgroup of the vector
flavour symmetry).  Eight of these gauge bosons acquire a mass through
the Higgs mechanism and the coloured excitations are lifted from the
low-energy spectrum.  There remain $10$ massless Nambu-Goldstone
excitations: a pseudo-scalar axial flavour octet of mesons, a scalar
superfluid boson associated with the broken U$(1)_{\text{B}}$ baryon
number generator, and a pseudo-scalar $\eta'$ boson associated with
broken axial U$(1)_{\text{A}}$ generator.  There remains one massless
gauge boson that is a mixture of the original photon and one of the
gluons~\cite{Alford:1998mk,Alford:1999pb}.  With respect to this
``rotated electromagnetism'' U$(1)_{\tilde{Q}}$ the CFL state remains
neutral~\cite{Alford:2002kj}.

The degeneracy of the vacuum manifold is lifted by the inclusion of a
non-zero strange quark mass $m_s$.  In the absence of instanton
effects and other quark masses, the ground state is not near to the
parity even CFL state~(\ref{eq:diquark}), but rather, is a kaon
rotated state CFLK$^{0}$.  As $m_s \rightarrow 0$ this state
approaches a state on the vacuum manifold that is a pure kaon rotation
of the parity even CFL~(\ref{eq:diquark}).

Even in the absence of quark masses, the vacuum manifold degeneracy is
partially lifted by the anomalous breaking of the U$(1)_{\text{A}}$
axial symmetry which we have neglected: Instanton effects tend to
disfavour kaon condensation by favouring parity even states, and thus
delay the onset of the CFLK$^{0}$ until $m_s$ reaches a critical value
(possibly excluding it).  The effects of anomaly and instanton
contributions have been well
studied~\cite{Rapp:1998zu,Rapp:1999qa,Schafer:1999fe,Manuel:2000wm,Son:2001jm,Schafer:2002ty}
and play an important quantitative role in the phase structure of QCD.
Non-zero up and down quark masses also tend to disfavour kaon
condensation. 

For the purposes of this paper, we shall neglect both the effects of
instantons, and the effects of finite up and down quark masses.  This
will ensure that kaon condensation occurs for arbitrarily small $m_s$.
Both of these effects open the possibility of a much richer phase
structure, including condensation of other mesons (see for
example~\cite{Kaplan:2001qk,Kryjevski:2004cw}).  Future analyses
should take these numerically important effects into account, both in
the effective theory and in the NJL model.

The primary source of for kaon condensation is the finite strange
quark mass.  To lowest-order, this behaves as a chemical
potential~\cite{Schafer:2000ew,Bedaque:2001je,Kaplan:2001qk} (see
(\ref{eq:ms=muy1}) and (\ref{eq:ms=muy2})).  In this paper, we also
consider the addition of a hypercharge chemical potential as this
removes many complications associated with masses and leads to a very
clean demonstration of kaon condensation.
\begin{figure}[b]
  {\small
    \psfrag{0}{$0$}
    \psfrag{20}{$20$}
    \psfrag{40}{$40$}
    \psfrag{60}{$60$}
    \psfrag{80}{$80$}
    \psfrag{450}{$450$}
    \psfrag{500}{$500$}
    \psfrag{550}{$550$}
    \psfrag{p (MeV) muY=0}{$\abs{p}$ (MeV) $\qquad\mu_Y = 0$}
    \psfrag{p (MeV) muY=muYc/2}{$\abs{p}$ (MeV) $\qquad\mu_Y = \mu_Y^c/2$}
    \psfrag{p (MeV) muY=muYc}{$\abs{p}$ (MeV) $\qquad\mu_Y = \mu_Y^c$}
    \psfrag{Ep (MeV)}{$E_{p}$ (MeV)}
    \includegraphics{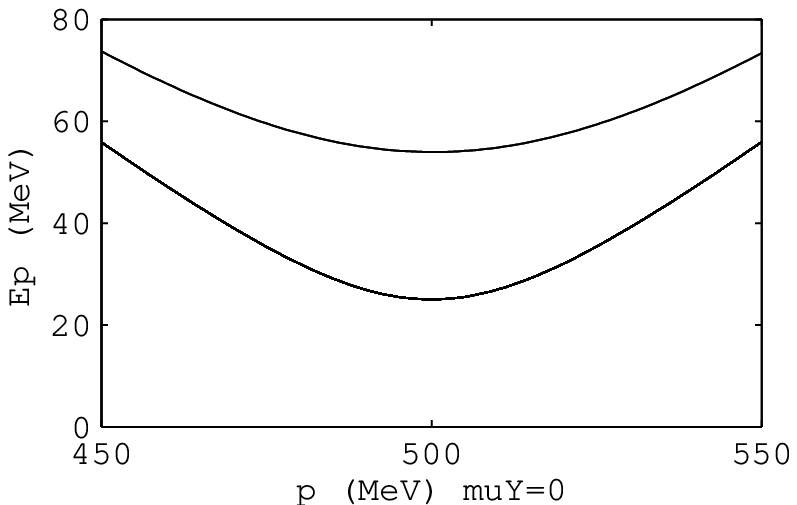}
    \includegraphics{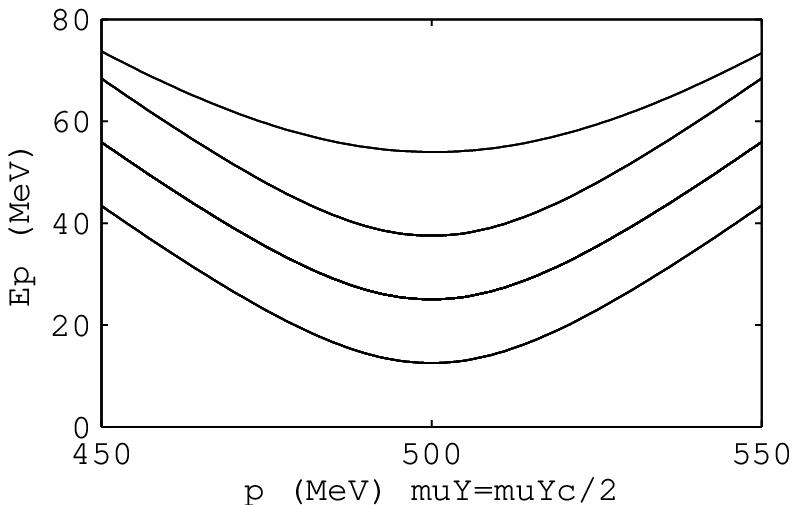}
    \includegraphics{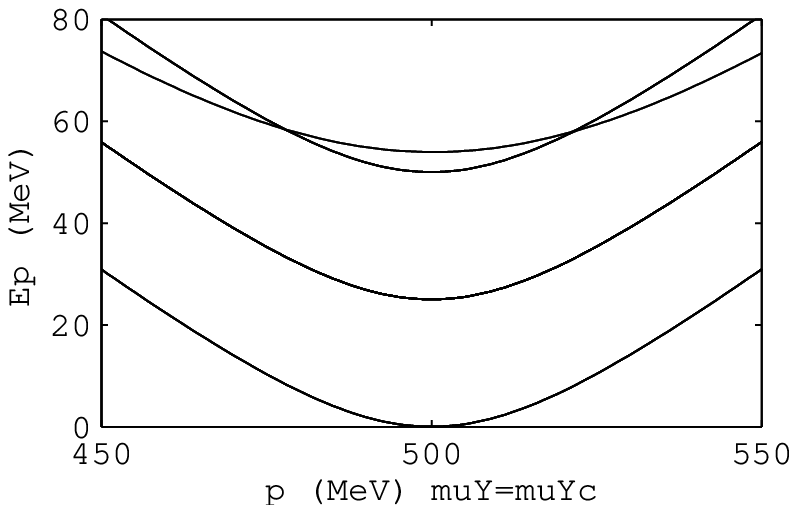}
    \caption{\label{fig:YDispersions} Lowest lying quasiparticle
      dispersions about the Fermi momentum $p_F = \mu_q = 500$ MeV for the
      CFL phase with different values of the hypercharge chemical.
      All dispersions have left-right degeneracy: we now consider the
      colour-flavour degeneracy.  In the top plot $\mu_Y = 0$, and the
      lowest dispersion has an eight-fold degeneracy and a gap of
      $\Delta_0 = 25$ MeV.  The upper band contains a single
      quasiparticle pairing (ru,gd,bs) with a gap of
      $4\Delta_6+2\Delta_3 = 54$ MeV.  In the middle plot, $\mu_Y =
      \mu_Y^c/2 = 12.5$ MeV, and (rs,bu) and (gs,bd) pairs are
      shifting as indicated in Table~\ref{tab:muShift}.  In the last
      plot, two pairs have become gapless marking the CFL/gCFL
      transition.}
  }
\end{figure}
\begin{figure}[b]
  {\small
    \psfrag{0}{$0$}
    \psfrag{20}{$20$}
    \psfrag{40}{$40$}
    \psfrag{60}{$60$}
    \psfrag{80}{$80$}
    \psfrag{450}{$450$}
    \psfrag{500}{$500$}
    \psfrag{550}{$550$}
    \psfrag{p (MeV) muY=muYc/2}{$\abs{p}$ (MeV) $\qquad\mu_Y = \mu_Y^c/2$}
    \psfrag{p (MeV) muY=1.20muYc}{$\abs{p}$ (MeV) $\qquad\mu_Y = 1.20\mu_Y^c$}
    \psfrag{Ep (MeV)}{$E_{p}$ (MeV)}
    \includegraphics{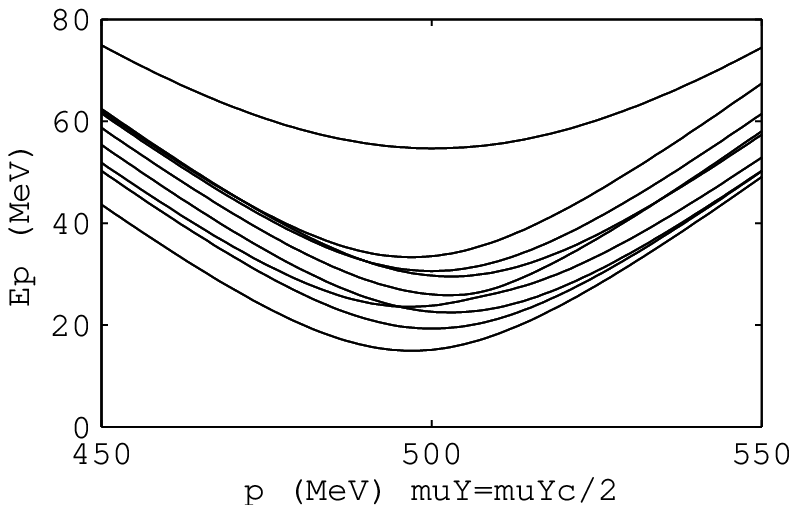}
    \includegraphics{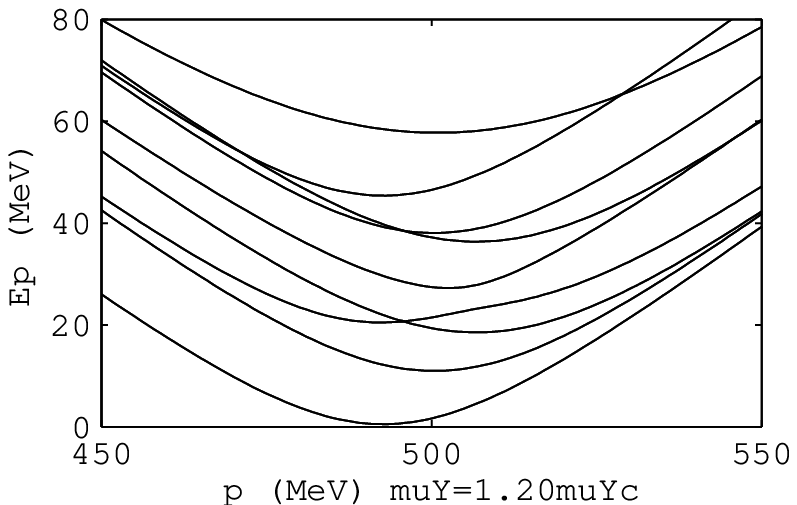}
    \caption{\label{fig:K0YDispersions} Lowest lying quasiparticle
      dispersions about the Fermi momentum $p_F = \mu_q = 500$ MeV for
      the CFLK$^0$ phase with different values of the hypercharge
      chemical. (The $\mu_Y=0$ dispersions are the same as in the top
      of Figure~\ref{fig:YDispersions}.)  Again, all dispersion have a
      left-right degeneracy.  In the top plot at $\mu_Y = \mu_Y^c/2 =
      12.5$ MeV, the eight-fold degenerate lowest band has split into
      eight independent dispersions.  To leading order in the
      perturbation, the splitting is described by
      Table~\ref{tab:muShift}, but the lack of degeneracy indicates
      that there are also higher order effects.  The lower plot at at
      $\mu_Y \approx 1.20\mu_Y^c \approx 30$ MeV is close to the
      CFLK$^0$/gCFLK$^0$ transition.  The gapless band now contains
      only a single mode and is charged.}}
\end{figure}
\section{Self-Consistent Solutions}
\label{sec:self-cons-solut}
We consider four qualitatively different phases: Two are
self-consistent mean-field solutions to the NJL model with a finite
hypercharge chemical potential parameter $\mu_Y$; the other two are
self-consistent mean-field solutions to the NJL model with a finite
strange-quark mass parameter $m_s$.  In each of these cases, one
solution corresponds to a parity even CFL phase and the other
corresponds to a kaon-condensed CFLK$^{0}$ phase.  Our normalizations
and a complete description of the model are presented in
Section~\ref{sec:njl-model}.  A full description of all the parameters
required to describe these phases along with some typical values is
presented in Appendix~\ref{sec:full-param}.
\begin{table}[h]
  \begin{center}
    \begin{tabular}{l|c|c|c|c|c|c|c|c|c|}
      &ru & gd & bs & rd & gu & rs & bu & gs & bd\\
      \hline
      CFL   & 0 & 0 & 0 & 0 & 0 & $-1$ & $+1$ & $-1$ & $+1$\\
      \hline
      CFLK$^0$ & 0 & $+\tfrac{1}{2}$ & $-\tfrac{1}{2}$ &
      0 & $+\tfrac{1}{2}$ &
      $-1$ & $+\tfrac{1}{2}$ &
      $-\tfrac{1}{2}$ & $+\tfrac{1}{2}$\\
      \hline
    \end{tabular}
    \caption{\label{tab:muShift} Leading order shifts in the chemical
      potentials of the various quarks in the CFL and CFLK$^0$ states
      in the presence of a hypercharge chemical potential shift
      $\mu_Y$.  This follows directly from~(\ref{eq:muColour}).}
  \end{center}
\end{table}  
\subsection{Finite Hypercharge Chemical Potential}
\label{sec:finite-hyperch-chem}
The CFL phase in the presence of a hypercharge chemical potential
corresponds to the fully gapped CFL phase discussed
in~\cite{Alford:2003fq}.  Here one models the effects of the strange
quark through its shift on the Fermi surface $p_F\approx \mu_q$ of the
strange quarks.  This can be seen by expanding the free-quark
dispersion
\begin{equation}
  \label{eq:ms=muy1}
  \sqrt{p^2+M_s^2} \approx \abs{p}+\frac{M_s^2}{2\mu_q}+\cdots
\end{equation}
or, more carefully, by integrating out the antiparticles to formulate
the High-Density Effective Theory. (See for
example~\cite{hong:1998tn,hong:1999ru,Beane:2000ms,Schafer:2001za}.)
These leading order effects are equivalent to adding a hypercharge
chemical potential of magnitude
\begin{equation}
  \label{eq:ms=muy2}
  \mu_Y = \frac{M_s^2}{2\mu_q}.
\end{equation}
and a baryon chemical potential shift of
\begin{equation}
  \label{eq:deltaMuB}
  \delta\mu_B = -\frac{M_s^2}{\mu_q}.
\end{equation}
We consider only the effect of the hypercharge chemical potential
here, holding $\mu_B$ fixed.  Note that the relevant parameters
are $M_s$ and $\mu_q$ rather than $m_s$ and $\mu_s = \mu_B/3$.  $M_s$
is the constituent quark mass that appears in the dispersion relation
whereas $m_s$ is the bare quark mass parameter; likewise, $\mu_q$ is
the corrected quark chemical potential that determines the Fermi
surface whereas $\mu_s = \mu_B/3$ is the bare baryon chemical
potential.  (These distinctions are important because our model takes
into account self-energy corrections.)
\begin{figure}[t]
  {\small
    \psfrag{minE/Delta0}
    {$\min_{p}E_{p}/\Delta_0$}
    \psfrag{muY/muYc}{$\mu_Y/\mu_Y^c$}
    \psfrag{0}{$0$}
    \psfrag{1}{$1$}
    \psfrag{0.2}{$0.2$}
    \psfrag{0.4}{$0.4$}
    \psfrag{0.5}{$0.5$}
    \psfrag{0.6}{$0.6$}
    \psfrag{0.8}{$0.8$}
    \psfrag{1.0}{$1.0$}
    \psfrag{1.2}{$1.2$}
    \includegraphics{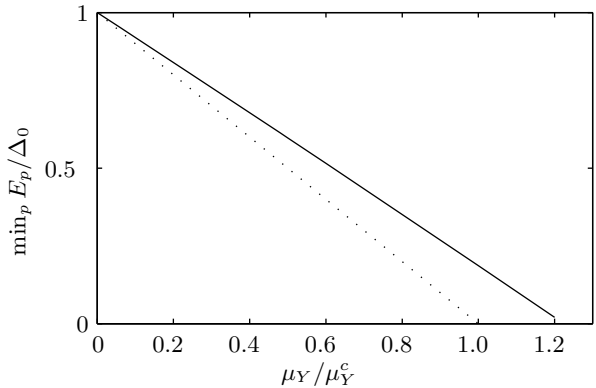}
    \caption{\label{fig:GapY} Physical gap of the lowest lying
      excitation as a function of the hypercharge chemical potential.  The
      dotted line corresponds to the CFL phase: the phase transition to
      the gCFL occurs at $\mu_Y = \mu_Y^c$ where the gap vanishes.  The
      solid line corresponds to the CFLK$^0$ state.  The transition to a
      gapless phase is delayed by a factor of $1.22$.}}
\end{figure}

The CFL phase responds in a trivial manner to a hypercharge chemical
potential: the quasiparticle dispersions shift such that the physical
gap in the spectrum becomes smaller; none of the other physical
properties change.  In particular, as the hypercharge chemical
potential increases, the coloured chemical potential $\mu_8=-\mu_Y$
decreases to maintain neutrality.  The values of all of the gap
parameters, the self-energy corrections, the densities and the
thermodynamic potential remain unchanged until the physical gap in the
spectrum vanishes.  (The apparent change in the magnitude of the gap
parameters in the first figure of~\cite{Alford:2003fq} is due to the
shift in the baryon chemical~(\ref{eq:deltaMuB}) which occurs if one
uses the strange quark chemical potential shift $\mu_s$ rather than a
hypercharge shift $\mu_Y$.)  This is a consequence of the $\tilde{Q}$
neutrality of the CFL state~\cite{Rajagopal:2000ff}.  In particular,
the electric chemical potential remains zero $\mu_e=0$ and the state
remains an insulator until the onset of the gapless modes.  The same
phenomena has also been noticed in the two-flavour
case~\cite{Bedaque:1999nu,Alford:2000sx,Kundu:2001tt}.
\begin{figure}[t]
  {\small
    \psfrag{minE/Delta0}
    {$\min_{p}E_{p}/\Delta_0$}
    \psfrag{ms2/2/mu/muYc}{$M_s^2/(2\mu_q\mu_Y^c)$}
    \psfrag{0}{$0$}
    \psfrag{1}{$1$}
    \psfrag{0.2}{$0.2$}
    \psfrag{0.4}{$0.4$}
    \psfrag{0.5}{$0.5$}
    \psfrag{0.6}{$0.6$}
    \psfrag{0.8}{$0.8$}
    \psfrag{1.0}{$1.0$}
    \psfrag{1.2}{$1.2$}
    \includegraphics{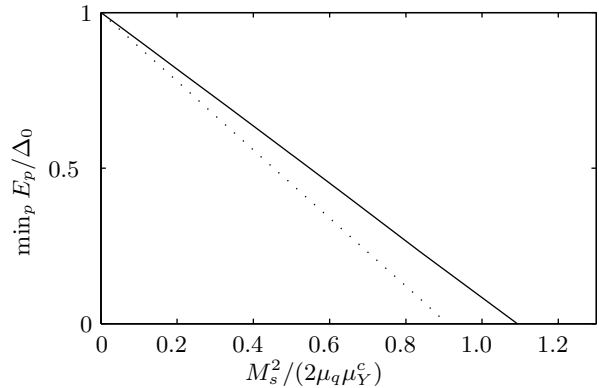}
    \caption{\label{fig:GapK0} Physical gap of the lowest lying
      excitation as a function of the strange quark mass.  The dotted
      line corresponds to the CFL phase and the solid line corresponds
      to the CFLK$^0$ phase.  We have normalized the axes in terms of
      $\mu_Y^c=\Delta_0$ for comparison with the hypercharge chemical
      potential case.  The CFL/gCFL transition occurs at a slightly
      smaller value of $M_s^2/\mu_q \approx 45.4$ MeV than the value
      of $46.8$ MeV
      in~\cite{Alford:2003fq,Alford:2004hz,Fukushima:2004zq}.  This is
      due to the effects of the other parameters on the quasiparticle
      dispersion relations.  We note that, as with $\mu_Y$, the
      transition from the CFLK$^0$ to a gapless phase is delayed
      relative to the CFL/gCFL transition, but by a slightly reduced
      factor of $1.2$.  This is in qualitative agreement but
      quantitative disagreement with the factor of $4/3$ found
      in~\cite{Kryjevski:2004kt}.  The is most-likely the result of
      our fully self-consistent treatment of the condensate
      parameters.}}
\end{figure}
As such, we can analytically identify the phase transition to the gCFL
phase which occurs for the critical chemical potential
\begin{equation}
  \label{eq:muYc}
  \mu_Y^c = \Delta_0
\end{equation}
where $\Delta_0 = \Delta_3-\Delta_6$ is the physical gap in the
spectrum in the absence of any perturbations.  Throughout this paper
we use parameters arbitrarily chosen so that $\mu_Y^c = \Delta_0 =
25$~MeV to correspond with the parameter values
in~\cite{Alford:2003fq,Alford:2004hz,Fukushima:2004zq}.  We show typical
quasiparticle dispersion relations for this state in
Figure~\ref{fig:YDispersions}.  

The splitting of the dispersions can also be easily understood from
the charge neutrality condition~(\ref{eq:muColour}) and the leading
order effects are summarized in Table~\ref{tab:muShift}.  After
setting $\mu_8=-\mu_Y$, the chemical potentials for the rs and gs
quarks shift by $-\mu_Y$ whereas for the bu and bd quarks it shifts by
$+\mu_Y$.  Thus, the (gs,bd) and (rs,bu) pairs are the first to become
gapless.
\begin{figure}[t]
  {\small
    \psfrag{0}{$0$}
    \psfrag{20}{$20$}
    \psfrag{40}{$40$}
    \psfrag{60}{$60$}
    \psfrag{80}{$80$}
    \psfrag{450}{$450$}
    \psfrag{500}{$500$}
    \psfrag{550}{$550$}
    \psfrag{p (MeV) muY=muYc/2}{$\abs{p}$ (MeV) $\qquad M_s^2/(2\mu_q) = 0.50\mu_Y^c$}
    \psfrag{p (MeV) muY=muYc}{$\abs{p}$ (MeV) $\qquad M_s^2/(2\mu_q) = 0.83\mu_Y^c$}
    \psfrag{Ep (MeV)}{$E_{p}$ (MeV)}
    \includegraphics{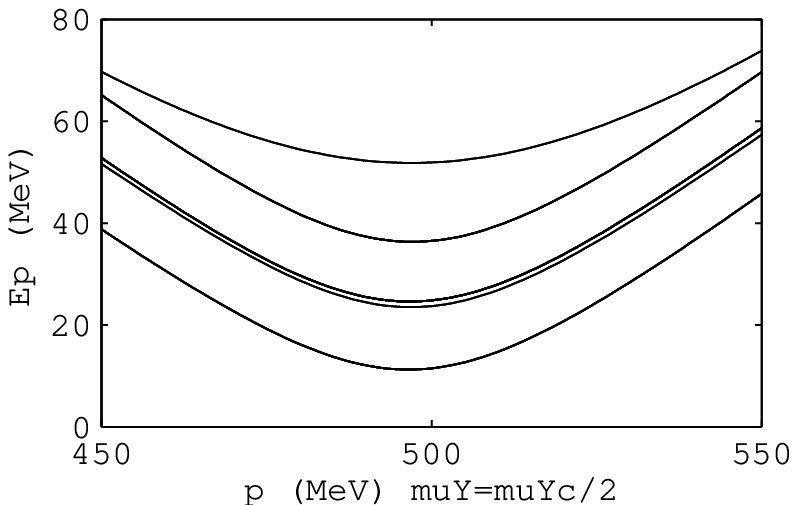}
    \includegraphics{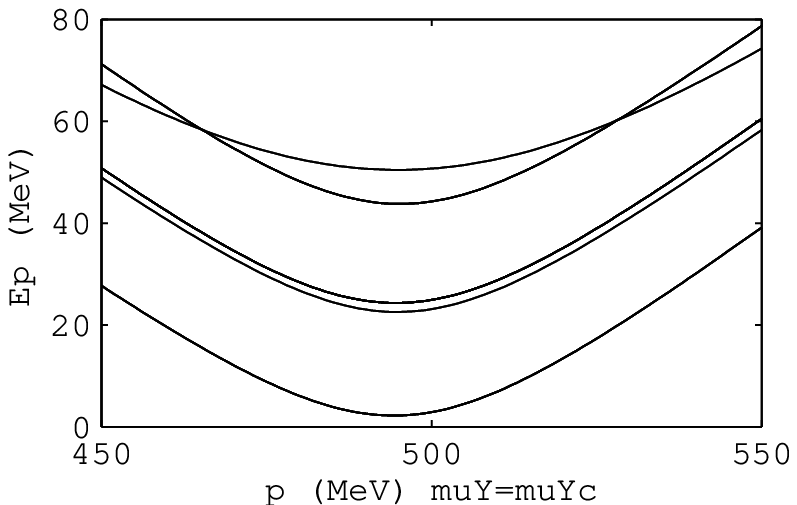}
    \caption{\label{fig:MSDispersions} Lowest lying quasiparticle
      dispersion relationships about the Fermi momentum $p_F = \mu_q =
      500$ MeV for the CFL phase with two different values of the
      strange quark mass. (The $M_s = 0$ dispersions are the same as in
      the top of Figure~\ref{fig:YDispersions}.) Qualitatively this has
      the same structure as Figure~\ref{fig:YDispersions} except that
      middle dispersion is now split by higher order mass effects.}}
\end{figure}

The kaon-condensed hypercharge state is more complicated.  One can
again use the appropriate charge neutrality
conditions~(\ref{eq:muColour}) to estimate how the quarks will be
affected by $\mu_Y$, but the na\"\i{}ve results hold only to
lowest-order.  In particular, the condensates of the CFLK$^0$ state
also vary as $\mu_Y$ increases (see Table~\ref{tab:CFLYK0}).  These
higher order effects break all the degeneracy between the quark
species and Figure~\ref{fig:K0YDispersions} has nine independent
dispersions.

We shall compare the thermodynamic potentials of these two states
later (see Figures~\ref{fig:fpiY} and~\ref{fig:fpims}), but we point
out here that the transition to a gapless colour-flavour--locked state
with kaon condensation (gCFLK$^0$) occurs at a larger hypercharge
chemical potential than the CFL/gCFL transition.  This can be most
easily seen in Figure~\ref{fig:GapY}.  This is in qualitative
agreement with~\cite{Kryjevski:2004kt} and~\cite{Kryjevski:2004jw},
but in quantitative disagreement.

In the CFL/gCFL transition, two modes become gapless simultaneously:
the lower branches of the (rs,bu) and (gs,bd) pairs.  One of these
modes is electrically neutral (gs,bd) and it crosses the zero-energy
axis giving rise to a ``breach'' in the spectrum.  The other mode is
electrically charged: as soon as in crosses, the electric chemical
potential must rise to enforce neutrality.  The state now contains
gapless charged excitations and becomes a conductor.  The result is
that the the neutral gapless mode has two linear dispersions while the
charged gapless mode has a virtually quadratic dispersion when
electric neutrality is enforced.  (This was discovered
in~\cite{Alford:2003fq} and is explained in detail
in~\cite{Alford:2004hz}.)
\begin{figure}[t]
  {\small
    \psfrag{0}{$0$}
    \psfrag{20}{$20$}
    \psfrag{40}{$40$}
    \psfrag{60}{$60$}
    \psfrag{80}{$80$}
    \psfrag{450}{$450$}
    \psfrag{500}{$500$}
    \psfrag{550}{$550$}
    \psfrag{p (MeV) muY=muYc/2}{$\abs{p}$ (MeV) $\qquad M_s^2/(2\mu_q) = 0.50\mu_Y^c$}
    \psfrag{p (MeV) muY=muYc}{$\abs{p}$ (MeV) $\qquad M_s^2/(2\mu_q) = 0.84\;\mu_Y^c$}
    \psfrag{Ep (MeV)}{$E_{p}$ (MeV)}
    \includegraphics{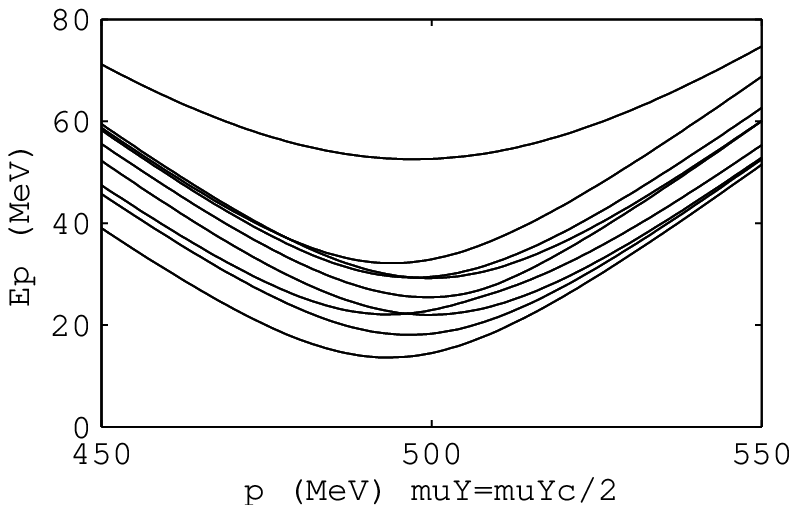}
    \includegraphics{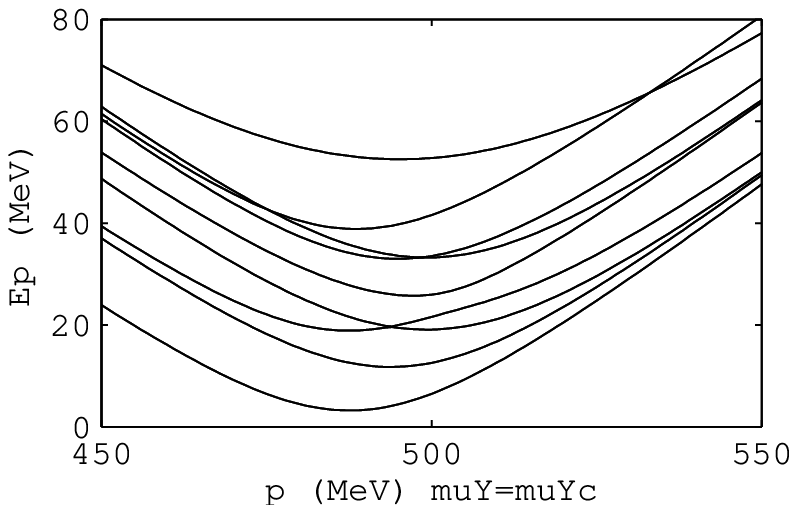}
    \caption{\label{fig:MSK0Dispersions} Lowest lying quasiparticle
      dispersion relationships about the Fermi momentum $p_F = \mu_q =
      500$ MeV for the CFLK$^0$ phase with two different values of the
      strange quark mass. (The $M_s=0$ dispersions are the same as in
      the top of Figure~\ref{fig:YDispersions}.)  Qualitatively this has
      the same structure as Figure~\ref{fig:K0YDispersions}.}}
\end{figure}

In the CFLK$^0$/gCFLK$^0$ transition, a single charged mode becomes
gapless.\footnote{This mode pairs rs, gu, and bu quarks in quite a
  non-trivial manner.  In the CFL, the quasiparticles form a nice
  block-diagonal structure in which the quarks exhibit definite
  pairing between two species.  In the CFLK0, the block structure is
  more complicated and the pairing cannot be simply described: the
  lowest lying quasiparticle is a linear combination of the
  three rs, gu, and bu quark.}  Thus, immediately beyond the transition,
the corresponding gCFLK$^0$ state will also be a conductor but there
will be a single charged gapless mode with almost quadratic
dispersion.  Additional modes will continue to lower until either more
modes become gapless, or a first order phase transition to a competing
phase occurs.
\subsection{Finite Strange Quark Mass}
The second pair of CFL/CFLK$^0$ states that we consider are
self-consistent solutions to the gap equation in the presence of a
finite strange quark mass.  Qualitatively we expect to see similar
features to the states at finite hypercharge chemical potential and
indeed we do as shown in Figures~\ref{fig:MSDispersions}
and~\ref{fig:MSK0Dispersions}.

Quantitatively, we notice a few differences with previous analyses
concerning the locations of the phase transitions to gapless states.
Our parameters have been chosen to match the parameters
in~\cite{Alford:2003fq,Alford:2004hz,Fukushima:2004zq}.  They find
that the gCFL/CFL transition occurs at $M_s^2/\mu =46.8 $ MeV, but the
CFL/gCFL transition happens noticeably earlier with our model at
$M_s^2/\mu = 43.9$ MeV.  This is due to a corresponding six-percent
reduction in the condensate parameters and represents the effects of
performing a fully self-consistent calculation.

Another difference concerns the appearance of gapless modes in the
CFLK$^0$ state (see Fig.~\ref{fig:GapK0}).  This transition occurs at
$M_s^2/\mu = 52.5$ MeV in our model---a factor of $1.2$ larger than
the CFL/gCFL transition.  This is some ten percent smaller than the
factor of $4/3$ derived in~\cite{Kryjevski:2004kt}.  This is likely
due to the more complicated condensate structure we consider and the
inclusion of self-energy corrections.
\section{NJL Model}\label{sec:njl-model} 
We base our analysis on the following Hamiltonian for the NJL model
\begin{equation}
  \label{eq:H}
  H = \int\frac{\d^3{\vect{p}}}{(2\pi)^3}
  \psi^\dagger_{\vect{p}}\left(
    \vect{\mat{\alpha}}\cdot\vect{p}
    -\mat{\mu}
    +\mat{\gamma}_0\mat{M}
  \right)\psi_{\vect{p}}
  +H_{\text{int}}.
\end{equation}
Here we consider $9$ species of quarks = $3$ colours $\times$ $3$
flavours: Including the relativistic structure, there are $36$ quark
operators in the vector $\psi$.  The matrices $\mat{\mu}$ and
$\mat{M}$ are the quark chemical potentials and masses respectively.

We take the interaction to be a four-fermion contact interaction with
the quantum numbers of single gluon exchange:\footnote{Here the
  matrices $\mat{\lambda}^A$ are the eight $3\times 3$ Gell-Mann
  matrices and the $\mat{\gamma}^{\mu}$ are the Dirac matrices which
  we take in the chiral basis.  Our normalizations and conventions are
  \begin{align*}
    \tr[\mat{\lambda}^A\mat{\lambda}^B] &= \tfrac{1}{2}\delta^{AB},\\
    \mat{\gamma}_5 &=
    i\mat{\gamma}_0\mat{\gamma}_1\mat{\gamma}_2\mat{\gamma}_3 = \begin{pmatrix}
      -\mat{1} & \mat{0}\\
      \mat{0} & \mat{1}
    \end{pmatrix},\\
    \mat{\gamma}_C &= i\mat{\gamma}_2\mat{\gamma}_0.
  \end{align*}
  We also use natural units where $c=\hbar=k_B = 1$.}
\begin{equation}
  H_{\text{int}} = 
  g\int
  (\bar{\psi}\mat{\gamma}^{\mu}\mat{\lambda}^A\psi)
  (\bar{\psi}\mat{\gamma}_{\mu}\mat{\lambda}^A\psi).
\end{equation}
The Gell-Mann matrices act on the colour space and the flavour
structure is diagonal.  We point out that this form of NJL interaction
has the desirable feature of explicitly breaking the independent
colour SU$(3)_{\text{CL}}$ left and SU$(3)_{\text{CR}}$ right
symmetries that some NJL models preserve.  This is important because
the condensation pattern~(\ref{eq:diquark}) does not explicitly link
left and right particles: Our model thus has the same continuous symmetries
as QCD, and the only complication to deal with is the gauging of the
single colour SU$(3)_{\text{C}}$ symmetry.

Our goal here is to provide a non-perturbative model to discuss the
qualitative features of QCD at finite densities.  We model the finite
density by working in the grand thermodynamic ensemble and introducing
a baryon chemical potential for all of the quarks:
\begin{equation}
  \mat{\mu} = \frac{\mu_{B}}{3}\mat{1}.
\end{equation}
With only this chemical potential and no quark masses, our model has
an U$(3)_{\text{L}}\otimes \text{U}(3)_{\text{R}}\otimes
\text{SU}(3)_{\text{C}}/\text{Z}_{3}$ continuous global symmetry in
which the left-handed quarks transform as
$(\bar{\mathbf{3}},\mathbf{1},\mathbf{3})$ and the right handed quarks
transform as $(\mathbf{1},\bar{\mathbf{3}},\mathbf{3})$.  In the
chiral basis we have explicitly\footnote{From this explicit
  representation we can see how the centres of the colour and
  flavours overlap giving rise to the Z$_{3}$ factor.}
\begin{equation}
  \label{eq:ExplicitSym}
  \begin{pmatrix}
    \psi_L\\
    \psi_R
  \end{pmatrix}
  \rightarrow 
  \begin{pmatrix}
    e^{-i\theta_L}\mat{F}_{L}^*\otimes\mat{C} & \mat{0}\\
    \mat{0} & e^{-i\theta_R}\mat{F}_{R}^*\otimes\mat{C}
  \end{pmatrix}
  \begin{pmatrix}
    \psi_L\\
    \psi_R
  \end{pmatrix}
\end{equation}
where $\mat{F}$ and $\mat{C}$ are SU$(3)$ matrices.  For an attractive
interaction, this NJL model exhibits the same symmetry breaking
pattern as QCD~(\ref{eq:SSB}) with a restored axial symmetry.  The
difference between this NJL model and QCD is that the NJL model
contains no gauge bosons.  Thus, there are $18$ broken generators
which correspond to massless Goldstone bosons, and none of these is
eaten.  To effectively model QCD, we must remove the extra coloured
Goldstone bosons.  At the mean-field level, this is done by imposing
gauge neutrality
conditions~\cite{Alford:2002kj,Kryjevski:2003cu,Gerhold:2003js}.  Once
the appropriate chemical potentials are introduced, the dependence on
the vacuum expectation values of the coloured Goldstone modes is
cancelled and the low-energy physics of the NJL model matches that of
QCD.

The usual NJL model has a local interaction, but this is not
renormalizable and needs regulation.  For the purposes of this paper,
we introduce a hard cutoff on each of the momenta $\Lambda_{\vect{p}}
= \theta\bigl(\Lambda-\norm{\vect{p}}\bigr)$ to mimic the effects of
asymptotic freedom at large momenta:
\begin{multline*}
  H_{\text{int}} = 
  \frac{g}{(2\pi)^9}\int
    \d^3{\vect{p}}
    \d^3{\vect{p}'}
    \d^3{\vect{q}}
    \d^3{\vect{q}'}\;
    \Lambda_{p}\Lambda_{p'}\Lambda_{q}\Lambda_{q'}\times\\
  \times\delta^{(3)}(\vect{p}-\vect{p}'+\vect{q}-\vect{q}')
  (\bar{\psi}_{\vect{p}}\mat{\gamma}^{\mu}\lambda^A\psi_{\vect{p}'})
  (\bar{\psi}_{\vect{q}}\mat{\gamma}_{\mu}\lambda^A\psi_{\vect{q}'}).
\end{multline*}
To study this model we perform a variational calculation by introducing
the quadratic Hamiltonian
\begin{equation}
  \label{eq:H0}
  H_0 = \int\frac{\d^3{\vect{p}}}{(2\pi)^3}\left(
    \psi^\dagger_{\vect{p}}\mat{E}(\vect{p})\psi_{\vect{p}}
    +
    \psi^T_{\vect{p}} \mat{B}\psi_{\vect{p}}
    +
    \psi^\dagger_{\vect{p}}\mat{B}^\dagger\psi^*_{\vect{p}}
  \right)
\end{equation}
where
\begin{equation}
  \mat{E}(\vect{p}) = \vect{\mat{\alpha}}\cdot\vect{p}
  -\mat{\mu}+\mat{\gamma}_0\mat{M}-\mat{A}
\end{equation}
and then computing the following upper
bound~\cite{feynman98:_statis_mechan} on the thermodynamic potential
$\Omega$ of the full system:
\begin{equation}
  \label{eq:Omega}
  \Omega \leq \Omega_0 + \braket{H-H_0}_0.
\end{equation}
$\Omega_0$ is the thermodynamic potential of the quadratic model and
the expectation value $\braket{}_0$ is the thermal average with
respect to the quadratic ensemble defined by $H_0$.  In
principle, the quadratic model is exactly solvable, thus the upper
bound can be computed.  One then varies the parameters $\mat{A}$ and
$\mat{B}$ to minimize this upper bound, obtaining a variational
approximation for the true ensemble.  In the zero-temperature limit,
this is equivalent to simply minimizing the expectation value of the
Hamiltonian over the set of all Gaussian states.

In practice, it is difficult to vary with respect to all possible
quadratic models since the space is of uncountable dimensionality.  In
this paper we restrict ourselves to minimizing over homogeneous and
isotropic systems.  This is equivalent to performing a fully
self-consistent mean-field analysis.  The condition for the right hand
side of~(\ref{eq:Omega}) to be stationary with respect to the
variational parameters is equivalent to the self-consistent gap
equation.

The microscopic analysis presented in this paper consists of choosing
reasonable parameterizations of $\mat{A}$ (which includes the chemical
potentials, masses and related corrections) and $\mat{B}$ (which
includes the gap parameters $\Delta$) that are closed under the
self-consistency condition, and numerically finding stationary points
of this system of equations.  (As $\mat{A}$ and $\mat{B}$ are
arbitrary $36\times 36$ matrices subject only to $\mat{A} =
\mat{A}^\dagger$ and $\mat{B} = -\mat{B}^T$, a full parametrization
consists of $2556$ parameters and was too costly for the present
author to consider.  However, the parametrization chosen is quite
natural and \emph{fully} closed.) Once the parameters $\mat{A}$ and
$\mat{B}$ are found, the properties of the ensemble can be computed by
diagonalizing the quadratic Hamiltonian.

As discussed in Section~\ref{sec:charge-neutrality}
and~\cite{Alford:2002kj}, we must impose the appropriate gauge charge
neutrality conditions.  This is done by introducing bare gauge
chemical potentials into the model and choosing these to ensure the
final solution is neutral.

To impose a charge neutrality condition, we instead vary $\mu_R$
(along with with the other parameters) to obtain a neutral solution
(again we note that the total charge and other correlations of the
state depend only on the corrected parameters $\mu_R$).  Once this
solution is found, $\delta\mu$ is computed and the required bare
chemical potential $\mu = \mu_R - \delta\mu$ determined.
Despite the fact that the self-energy corrections depend only on the
corrected parameters ($\mu_R$ etc.), the thermodynamic potential
depends on both the corrected and the bare parameters and so this last
step is important.

One must also be careful about which thermodynamic potential is used
to compare states when neutrality conditions are enforced as we are
no longer in the grand ensemble.  The differences between the
potentials of the relevant ensembles are proportional to terms of the
form $Q\mu$, however, so for neutrality conditions, $Q=0$, and
the thermodynamic potential may still be used to compare states.

\subsection{Numerical Techniques}
We sketch here the method used to calculate the thermodynamic
potential and perform the variational minimization.  First, we express
the Hamiltonian $H$ in the following simplified form
\begin{equation}
  H = \Psi^\dagger\mat{H}\Psi 
  + g\Psi^\dagger\mat{\Gamma}^\dagger\Psi\Psi^\dagger\mat{\Gamma}\Psi
\end{equation}
where $\mat{H}$ is a hermitian matrix.  In order to do this and
include the ``anomalous'' correlations $\braket{\psi\psi}$, we must use
an augmented ``Nambu-Gorkov'' spinor
\begin{equation}
  \label{eq:NambuGorkov}
  \Psi = \begin{pmatrix}
    \psi\\
    \psi^C
  \end{pmatrix},
\end{equation}
where $\psi^C$ is the charge conjugated spinor.  This doubling of the
degrees of freedom requires careful attention to avoid double
counting.

To simplify the presentation of our method in this section, we shall
ignore this complication and assume that $\Psi$ contains a single set
of operators with no duplicate degrees of freedom.  We also consider
only a single interaction term, and subsume the momentum structure
into the matrix structure.  Explicitly dealing with these
complications is straightforward and the details are presented
in~\cite{Forbes:2005}.\footnote{To derive the full equations, one
  introduces the augmented structure while imposing constraints on the
  matrices throughout the variation.  For example, one must ensure
  that $\mat{H} = -\mat{C}\mat{H}^{T}\mat{C}$ where $\Psi^{C} =
  \mat{C}\Psi$.  To derive the proper momentum structure, one simply attaches
  momentum indices: for homogeneous states all correlations have the
  form $\braket{\Psi_{\vect{p}}^\dagger\Psi_{\vect{q}}} \propto
  \delta^{3}(\vect{p}-\vect{q})$ and the momentum structure follows
  trivially.}

We now express the variational Hamiltonian as
\begin{equation}
  H_0 = \Psi^\dagger(\mat{H}+\mat{\Sigma})\Psi = \Psi^\dagger\mat{H}_0\Psi.
\end{equation}
The matrix of variational parameters $\mat{\Sigma}$ may be thought of
as the self-energy corrections.  All of the two-point correlations are
determined from the corrected Hamiltonian matrix $\mat{H}_0$, with the
``anomalous'' correlations being found off the diagonal:
\begin{align*}
  \braket{\Psi\Psi^\dagger} &= \mat{F}^{-}, &
  \braket{\Psi^*\Psi^T} &= [\mat{F}^{+}]^T, &
  \mat{F}^{\pm} &= \frac{1}{1+e^{\pm\beta \mat{H}_{0}}}.
\end{align*}
At finite temperatures, there is a one-to-one relationship between
$\mat{E}$ and the matrix $\mat{F}^{-} = \mat{1}-\mat{F}^{+}$.  Armed
with this result, the variational bound takes the explicit
form\footnote{For a fully self-consistent analysis, we must include
  the term $-g\braket{\Psi^\dagger\Psi^\dagger}\braket{\Psi\Psi}$.
  These correlations vanish in this simplified analysis, but are
  included when the full augmented structure~(\ref{eq:NambuGorkov}) is
  considered as discussed in~\cite{Forbes:2005}.  Note also that
  momentum integration is implicit in the matrix multiplication and
  traces.}
\begin{multline}
  \label{eq:OmegaBound1}
  \Omega \leq
  \frac{1}{\beta}\tr\ln[\mat{F}^{-}]
  -
  \tr[\mat{\Sigma}\mat{F}^{+}]
  +\\
  +
  g\left(
    \tr[\mat{\Gamma}^\dagger \mat{F}^{+}]
    \tr[\mat{\Gamma} \mat{F}^{+}]
    +
    \tr[\mat{\Gamma}^\dagger \mat{F}^{-}
    \mat{\Gamma}\mat{F}^{+}]
  \right).
\end{multline}
From this, one may find the stationary points by by differentiating
with respect to $\mat{F}^{+}$.  At finite temperature this is
formally equivalent to finding the stationary points by varying with respect to
$\mat{\Sigma}$.  Differentiating with respect to $\mat{\Sigma}$ is
complicated by the functional dependence and the result is not
expressible as a simple matrix equation.  The conditions
$\partial\Omega/\partial F^{+}_{ij} = 0$ yield the fully
self-consistent Schwinger-Dyson equations which may be expressed as:
\begin{equation}
  \label{eq:SchwingerDyson1}
  \mat{\Sigma}
  =
  g\left(
    \mat{\Gamma}^\dagger\tr[\mat{\Gamma} \mat{F}^{+}]
    +
    \mat{\Gamma}\tr[\mat{\Gamma}^\dagger \mat{F}^{+}]
    +
    \mat{\Gamma}^\dagger\mat{F}^{-}\mat{\Gamma}
    -
    \mat{\Gamma}\mat{F}^{+}\mat{\Gamma}^\dagger
  \right).
\end{equation}
In principle, one may derive analytic expressions for these equations,
but, in practise, the matrices are $72\times 72$ and computing
$\mat{F}^{\pm}$ analytically---even when many approximations are
made---is quite tedious.  Instead, we simply use these expressions
numerically.  The required diagonalization is then efficiently
performed using standard numerical linear algebra tools.  The traces
involved include momentum integrals, but for homogeneous states these
are one-dimensional and thus also quite efficient.  The biggest
challenge is to solve simultaneously the equations present
in~(\ref{eq:SchwingerDyson1}).  This is done by first projecting out
the the limited subspaces describe in Appendix~\ref{sec:full-param}
and then employing a multi-dimensional root-finder.

Since the search space is large ($\sim 45$ parameters for the CFLK$^0$
states), traditional root-finders are prohibitive because they
recompute the Jacobian at each step.  Here we use a modified Broyden
algorithm~\cite{Broyden:1965,DS:1983} to provide a secant-like update
to the Jacobian requiring far fewer function evaluations per step of
the algorithm.

In many cases, the Schwinger-Dyson equation~(\ref{eq:SchwingerDyson1})
converges through simple iteration.  With charge neutrality
constraints, this is often no longer the case, but the Broyden update
is sufficient to restore convergence.
\subsection{CFL at $m_s=0$.}
As an example, consider the parity even CFL state.  The
self-consistency conditions are fully closed when one includes four
variational parameters.  There are two gap parameters $\Delta_3$ and
$\Delta_6$ corresponding to the diquark condensate~(\ref{eq:diquark}),
one chemical potential correction $\delta\mu_B$ to the baryon chemical
potential and an induced off-diagonal chemical potential
$\mu_{\text{oct}}$.  The quadratic Hamiltonian~(\ref{eq:H0}) can thus
be expressed
\begin{equation*}
  H_0 = 
  \psi^\dagger\left(
    \vect{\mat{\alpha}}\cdot\vect{p}
    -\tfrac{1}{3}\mu_{B}
    -\mat{\delta\mu}
  \right)
  +
  \tfrac{1}{2}\psi^T \mat{\gamma}_C\mat{\gamma}_5 \mat{\Delta}
  \psi\; +\; \text{h.c.}
\end{equation*}
where
\begin{align}
  [\mat{\Delta}]^{\alpha a}{}_{\beta b} &= 
  \Delta_3\epsilon^{\alpha \beta k}\epsilon_{a b k}+
  \Delta_6(\delta^{\alpha}_{a}\delta^{\beta}_{b}+
  \delta^{\alpha}_{b}\delta^{\beta}_{a}),\\
  \mat{\delta\mu} &= \tfrac{1}{3}\delta\mu_B+\mat{\mu_{\text{oct}}},
\end{align}
and
\begin{equation*}
  [\mat{\mu}_{\text{oct}}]^{\alpha a}{}_{\beta b} =
  \mu_{\text{oct}}\left(
    \sum_{A=1}^{8}
    [\mat{\lambda}^A]_{\alpha a}
    [\mat{\lambda}^A]_{\beta b}
    -
    8
    [\mat{\lambda}^0]_{\alpha a}
    [\mat{\lambda}^0]_{\beta b}
  \right).
\end{equation*}
Most of this structure is all well-known and discussed many times in
the literature, however, there has been no mention of the parameter
$\mat{\mu_{\text{oct}}}$ because most analyses neglect the self-energy
corrections.

Neglecting the correction to the baryon chemical potential is
reasonable since it has little physical significance: it simply enters
as a Lagrange multiplier to establish a finite density.  As such, the
effective common quark chemical potential
\begin{equation}
  \mu_q = \tfrac{1}{3}\mu^{\text{eff}}_{B} =
  \tfrac{1}{3}\left(\mu_{B}+\delta\mu_B\right)
\end{equation}
is the relevant physical parameter defining the Fermi surface.  To
compare states in the grand ensemble, however, one must fix the bare
rather than the effective chemical potentials.  This is what we have
done in our calculations.  Numerically, we find that the corrections
$\delta\mu_B$ cause $\mu_q$ to vary by only a few percent as we vary
the perturbation parameters $\mu_Y$ and $m_s$.

There is no bare parameter corresponding to $\mu_{\text{oct}}$.  Thus,
it is spontaneously induced and should be treated on the same footing
as $\Delta$.  To see that such a parameter must exist, consider
changing to the ``octet'' basis using the augmented Gell-Mann matrices
\begin{equation}
  \tilde{\psi}_A = 2[\mat{\lambda}^A]_{\alpha a} \psi_{(\alpha a)}
\end{equation}
where $\mat{\lambda}^0 = \mat{1}/\sqrt{6}$.  In this basis, the
off-diagonal condensate becomes diagonal with one singlet parameter
$4\Delta_6+2\Delta_3$ and eight octet parameters $\Delta_6-\Delta_3$:
\begin{equation}
  \label{eq:DeltaOct}
  \tilde{\mat{\Delta}} = 
  \begin{pmatrix}
    4\Delta_6+2\Delta_3\\
    &\Delta_6-\Delta_3\\
    && \ddots \\
    &&&\Delta_6-\Delta_3
  \end{pmatrix}
\end{equation}
It is clear that in the CFL, the singlet channel decouples from the
octet channel: there is no symmetry relating these and the two gap
parameters are related by the numerical value of the coupling $g$.
This decoupling is also present in the chemical potential corrections.
One linear combination corresponds to the identity: this corrects the
baryon chemical potential $\delta\mu_B$.  The other is the induced
$\mu_{\text{oct}}$.

Numerically, we calibrate our model with this CFL solution.  In
particular, we chose our parameters to reproduce the results
of~\cite{Fukushima:2004zq}.  We use a hard cutoff at $\Lambda = 800$
MeV, and a coupling constant chosen so that, with an effective quark
chemical potential of $\mu_q = 500$ MeV one has a physical gap in the
spectrum of $\Delta_0 = \Delta_3-\Delta_6 = 25$ MeV.  This fixes the
following parameter values which we hold fixed for all of our
calculations:
\begin{subequations}
  \begin{align}
    \Lambda &= 800 \text{ MeV},\\
    g\Lambda^2 &=  1.385,\\
    \mu_B/3 &= 549.93 \text{ MeV}.
  \end{align}
\end{subequations}
With these parameters fixed, the fully self-consistent mean-field CFL
solution has the following variational parameters:
\begin{subequations}
  \begin{align*}
    \Delta_3 &= 25.6571 \text{ MeV},\\
    \Delta_6 &= 0.6571 \text{ MeV},\\
    \delta\mu_B/3 &= -49.93 \text{ MeV},\\
    \mu_{\text{oct}} &= -0.03133 \text{ MeV}.
  \end{align*}
\end{subequations}
As first noted in~\cite{Alford:1998mk}, and discussed
in~\cite{Alford:1999pa}, the parameter $\Delta_6$ is required to close
the gap equation, but is small because the sextet channel is
repulsive.  In weakly-coupled QCD, $\Delta_6$ is suppressed by an
extra factor of the coupling.  This effect is numerically captured in
the NJL model.  The parameter $\mu_{\text{oct}}$ is also required to
close the gap equation when the Hartree-Fock terms are included.  It
is also numerically suppressed.  Recent calculations often omit
$\Delta_6$ and $\mu_{\text{oct}}$: we see that this is numerically
justified.

The physical gap in the spectrum also defines the critical hypercharge
chemical potential for the CFL/gCFL transition~(\ref{eq:muYc}):
\begin{equation}
  \mu_Y^c = \Delta_0 = 25.00 \text{ MeV}.
\end{equation}
\subsection{CFL at $\mu_Y, m_s \neq 0$}
Once one introduces a strange quark mass, one must introduce
additional parameters.  A simple way to determine which parameters are
required is to add the mass, then compute the gap equation and see
which entries in the self-energy matrix are non-zero.  By doing this
for a variety of random values of the parameters, one can determine
the dimension of the subspace required to close the gap equation and
introduce the required parameters.

In the case of the CFL state with non-zero hypercharge chemical
potential, one only needs to introduce the parameters $\mu_Y$ and
$\mu_8$ to ensure gauge neutrality: As discussed in
Sec.~\ref{sec:finite-hyperch-chem}, none of the other parameters
change.  To go beyond the transition into the gCFL phase, however, or
to extend the results to non-zero temperature, one must introduce
additional parameters.  These include the perturbation $\mu_Y$, the
gauge chemical potentials $\mu_3$, $\mu_8$, and $\mu_e$ required to
enforce neutrality, as well as nine gap parameters $\Delta_{12}$,
$\Delta_{23}$, $\Delta_{13}$, $\Delta_{45}$, $\Delta_{67}$,
$\Delta_{89}$, $\Delta_{11}$, $\Delta_{22}$, and $\Delta_{33}$ that
fully parametrize the triplet and sextet diquark condensates.  (These
latter nine parameters correspond to the parameters $\phi_i$,
$\varphi_i$ and $\sigma_i$ defined in reference~\cite{Ruster:2004eg}.)
The additional parameters are chemical potentials similar to
$\mu_{\text{oct}}$ which are induced by the gap equations.  The full
set of parameters in discussed in Appendix~\ref{sec:full-param}.

Adding a strange quark mass is more complicated.  First of all, we
need to introduce additional Lorentz structure.  For homogeneous and
isotropic systems, there are eight possible relativistic structures:
\begin{align*}
  \mat{A} &= \mat{1}\otimes \delta\mat{\mu} 
  + \mat{\gamma}_5 \otimes \delta\mat{\mu}_5
  - \mat{\gamma}_0 \otimes \delta\mat{m}
  - \mat{\gamma}_0\mat{\gamma}_5\otimes \delta\mat{m}_5,\\
  \mat{B} &= \mat{\gamma}_C\mat{\gamma}_5\otimes\mat{\Delta} 
  + \mat{\gamma}_C\otimes \mat{\Delta}_5
  + \mat{\gamma}_0\mat{\gamma}_C\mat{\gamma}_5 \otimes \mat{\kappa}
  + \mat{\gamma}_0\mat{\gamma}_C \otimes\mat{\kappa}_5.
\end{align*}
Introducing quark masses requires one to introduce the additional
Lorentz structure $\mat{\kappa}$~\cite{Alford:1999pa} to close the gap
equations, but these are found to be small.  In total, one requires
about $20$ parameters to fully parametrize the CFL in the presence of
a strange quark mass (see Table~\ref{tab:CFLMs}).

With the inclusion of a bare quark mass $m_s$ one induces a chiral
condensate $\braket{\bar{\psi}\psi}$ which in turn generates a
correction to the quark mass.  The resulting parameter in
$\mat{H}_{0}$~(\ref{eq:H0}) is the constituent quark mass $M_s$ which
appears in the dispersion relationships for the quarks.  It is this
constituent quark mass that must be used when calculating the
effective chemical potential shift~(\ref{eq:ms=muy2}).  Generally the
constituent quark mass is quite a bit larger than the bare quark mass
parameter $m_s$.  For example, close to the phase transition, we have
$m_s \approx 83$ MeV while the constituent quark mass is $M_s \approx
150$ MeV (see Table~\ref{tab:CFLMs}).  We have checked that our
calculations are quantitatively consistent with the calculations
presented in~\cite{Buballa:2001gj} in this regard.

\subsection{CFLK$^0$}
Applying a kaon rotation to the CFL state breaks the parity of the
state, and mixes the parity even parameters $\mat{\mu}$, $\mat{m}$,
$\mat{\Delta}$ and $\mat{\kappa}$ with their parity odd counterparts
$\mat{\mu}_5$, $\mat{m}_5$, $\mat{\Delta}_5$ and $\mat{\kappa}_5$.
The full set of parameters and typical numerical values is presented
in Appendix~\ref{sec:full-param}.

\section{Low-Energy Effective Theory}
\label{sec:low-energy-effective}
To describe the low-energy physics of these models, we follow a well
established procedure: identify the low-energy degrees of freedom and
their transformation properties, identify the expansion parameters
(power counting scheme), write down the most general action consistent
with the symmetries and power counting, and determine the arbitrary
coefficients by matching to experiment or another theory.  In our
case, we will match onto the mean-field approximation of the NJL
model.  The resulting low-energy effective theory has been well
studied~\cite{hong:1999dk,Casalbuoni:1999wu}: we use this presentation
to establish our conventions, and to contrast the effective theory of
QCD with that of the microscopic NJL model.
\subsection{Degrees of Freedom}
The coset space in the NJL model is isomorphic to U(3)$\otimes$U(3).
This can be fully parametrized with two SU(3) matrices $\mat{X}$ and
$\mat{Y}$ and two physical phases $A$ and $V$ which one can physically
identify with the condensates:
\begin{subequations}
  \label{eq:XY}
  \begin{align}
    \sqrt{V^\dagger A}[\mat{X}]_{c\gamma} &\propto
    \epsilon_{abc}\epsilon_{\alpha\beta\gamma}
    \braket{\psi_L^{a\alpha}\psi_L^{b\beta}}, \\
    \sqrt{V^\dagger A^\dagger}[\mat{Y}]_{c\gamma} &\propto
    \epsilon_{abc}\epsilon_{\alpha\beta\gamma}
    \braket{\psi_R^{a\alpha}\psi_R^{b\beta}}.
  \end{align}
\end{subequations}
These thus transform as follows:
\begin{subequations}
  \label{eq:EffTrans}
  \begin{align}
    \mat{X} &\rightarrow \mat{F}_L\mat{X} \mat{C}^\dagger, \\
    \mat{Y} &\rightarrow \mat{F}_R\mat{Y} \mat{C}^\dagger,\\
    A &\rightarrow e^{2i(\theta_R-\theta_L)}A,\\
    V &\rightarrow e^{2i(\theta_R+\theta_L)}V.
  \end{align}
\end{subequations}
Note that the condensation pattern $\mat{X} = \mat{Y} = \mat{1}$,
$A=V=1$ is unbroken by the residual symmetry where $\mat{F}_L =
\mat{F}_R = \mat{C}$ and also by the $\text{Z}_2$ symmetries where
$\theta_L,\theta_R = \pm\pi$.  This is the reason for the extra factor
of two in the phases.  In QCD the degrees of freedom are similar, but
one must consider only colour singlet objects.  Thus, the low-energy
theory for QCD should include only the colour singlet combination
\begin{equation}
  \mat{\Sigma} = \mat{X}\mat{Y}^\dagger 
  \rightarrow \mat{F}_L \mat{\Sigma}\mat{F}_R^\dagger
\end{equation}
and the colour singlet phases $A$ and $V$.  Note also that these have
the following transformation properties under parity
\begin{subequations}
  \begin{align}
    \mat{X} &\leftrightarrow \mat{Y},&
    \mat{A} &\leftrightarrow \mat{A}^\dagger,&
    \mat{\Sigma} &\leftrightarrow \mat{\Sigma}^\dagger.
  \end{align}
\end{subequations}

The field content of the effective theories is thus:
\begin{description}
\item[$H$, $\eta'$:] Two singlet fields corresponding to the U$(1)$ phases of $A$ and
  $V$.  The field associated with $V$ is a scalar boson
  associated with the superfluid baryon number condensation.  We shall
  denote this field $H$.
  
  The field associated with $A$ is a pseudo-scalar boson associated
  with the axial baryon number symmetry and shall be identified with
  the $\eta'$ particle.  As discussed in
  Section~\ref{sec:colo-flav-lock}, the axial symmetry symmetry is
  anomalously broken in QCD and the $\eta'$ is not strictly massless
  due to instanton effects, but these are suppressed at high density.
  We ignore these effects.  Our NJL model thus contains no instanton
  vertex and our low-energy theory will contain no Wess-Zumino-Witten
  terms~\cite{Wess:1971yu,Witten:1983tx}.  It would be interesting to
  include both of these terms and repeat this calculation as these
  effects are likely not small~\cite{Schafer:2002ty}.
\item[$\pi^a$:] Eight pseudo-scalar mesons $\pi^a$ corresponding to the broken
  axial flavour generators.  As colour singlets these remain as
  propagating degrees of freedom in both QCD and NJL models.  These
  have the quantum numbers of pions, kaons and the eta and transform
  as an octet under the unbroken symmetry.
\item[$\phi^a$:] Eight scalar bosons $\phi^a$ corresponding to the broken
  coloured generators.  These are eaten by the gauge bosons in QCD and
  are removed from the low-energy theory.  This gives masses to eight
  of the gauge bosons and decouples them from low-energy physics.  In
  the NJL model these bosons still remain as low-energy degrees of
  freedom, but decouple from the colour singlet physics when one
  properly enforces colour neutrality.
\end{description}
There are additional fields and effects that should be considered as
part of a complete low-energy theory, but that we neglect:
\begin{enumerate}
\item The appropriately ``rotated electromagnetic field'' associated
  with the unbroken U$(1)_{\tilde{Q}}$ symmetry remains massless.
  Both the CFL and CFLK$^0$ states remain neutral with respect to this
  field, however, and we do not explicitly include it in our
  formulation.
\item The leptons are not strictly massless, but the electron and muon
  are light enough to consider in the low-energy physics.  In
  particular, they contribute to the charge density in the presence of
  an electric chemical potential and at finite temperature.  In this
  paper, leptonic excitations play no role since we consider only
  $T=0$ and both CFL and CFLK$^0$ quark matter is electrically neutral
  for $\mu_e=0$.  The leptons play an implicit role in fixing
  $\mu_{\tilde{Q}}$ such that $\mu_e = 0$ in both insulating phases.
\end{enumerate}
To be explicit, we relate all of the dimensional physical fields $H$,
$\eta'$, $\phi^a$ and $\pi^a$ to the phase angles through their decay
constants: $H = f_{H} \tilde{H}$, $\eta' = f_{\eta'} \tilde{\eta}'$,
$\phi^a = f_{\phi} \tilde{\phi}^a$, and $\pi^a = f_{\pi}
\tilde{\pi}^a$.  The two U$(1)$ phases angles have a slightly
different normalization because of the normalization of the
generators.  This normalization is chosen to match the kinetic terms
in the original theory and matches~\cite{Son:1999cm,Son:2000tu}:
$\tilde{\eta}'=\sqrt{6}(\theta_R-\theta_L)$, and
$\tilde{H}'=\sqrt{6}(\theta_R+\theta_L)$.  The realization of these
transformations in the microscopic theory is
\begin{equation}
  \label{eq:PsiTransform}
  \psi \rightarrow \exp\left\{
    i\left[
      \frac{-\tilde{H}\mat{1}-\tilde{\eta}'\mat{\gamma}_5}{2\sqrt{6}}
      +\tilde{\phi}^a \mat{r}^{a}
      +\tilde{\pi}^a \mat{f}^{a}_{A}
    \right]
  \right\}\psi
\end{equation}
where
\begin{subequations}
  \begin{align}
    \mat{f}^a_{R,L} &= (\mat{1}\pm\mat{\gamma}_5)\otimes(-\mat{\lambda}_a^*)\otimes\mat{1}/2,\\
    \mat{c}^{a} &= \mat{1}\otimes\mat{1}\otimes\mat{\lambda}_a,\\
    \mat{f}^a_A &= \mat{f}^a_R-\mat{f}^a_L 
    = \mat{\gamma}_5\otimes(-\mat{\lambda}_a^*)\otimes\mat{1},\\
    \mat{f}^a_V &= \mat{f}^a_R+\mat{f}^a_L 
    = \mat{1}\otimes(-\mat{\lambda}_a^*)\otimes\mat{1},\\
    \mat{r}^{a} &= (\mat{f}^{a}_{V}-\mat{c}^{a})/2 %,\\
  \end{align}
\end{subequations}
and the corresponding realization in the effective theory is
\begin{subequations}
  \label{eq:EffTransform}
  \begin{align}
    \mat{X} &=
    \exp\left\{-i\tilde{\pi}^a\mat{\lambda}_a\right\}
    \exp\left\{i\tilde{\phi}^a\mat{\lambda}_a\right\},\\
    \mat{Y} &=  
    \exp\left\{i\tilde{\pi}^a\mat{\lambda}_a\right\}
    \exp\left\{i\tilde{\phi}^a\mat{\lambda}_a\right\},\\
    A &= \exp\left\{2i\tilde{\eta}'/\sqrt{6}\right\},\\
    V &= \exp\left\{2i\tilde{H}/\sqrt{6}\right\},\\
    \mat{\Sigma} &= \exp\left\{-2i
      \tilde{\pi}^a\mat{\lambda}_a
    \right\}.
  \end{align}
\end{subequations}
\subsection{Power Counting}
In addition to $\Lambda_{\text{QCD}}$ which separates the three light
quarks from the heavy quarks, there are two primary scales in
high-density QCD: the quark chemical potential $\mu_q$ and the gap
$\Delta$.  In the NJL model there is also a cutoff and the coupling
constant: these are related by the gap equation when one holds $\mu$
and $\Delta$ fixed and the qualitative physics is not extremely
sensitive to the remaining renormalization parameter.

Our low-energy theory is an expansion in the energy/momentum of the
Goldstone fields.  Thus, the expansion is in powers of the derivatives
with respect to the scales $\mu$ and $\Delta$.  In this paper, we
shall only consider leading order terms: Systematic expansions have
been discussed elsewhere (see for example~\cite{Schafer:2003jn}).
\subsection{Kinetic Terms}
To construct the low-energy theory we follow~\cite{Casalbuoni:1999wu}
and introduce coloured currents
\begin{align*}
  \mat{J}^\mu_{X} &= \mat{X}^\dagger \partial^\mu_{(v)} \mat{X}
  \rightarrow \mat{C} \mat{J}^\mu_{X}\mat{C}^\dagger,\\
  \mat{J}^\mu_{Y} &= \mat{Y}^\dagger \partial^\mu_{(v)} \mat{Y}
  \rightarrow \mat{C} \mat{J}^\mu_{Y}\mat{C}^\dagger,\\
  \mat{J}_{\pm}^{\mu} &= \mat{J}_{X}^{\mu} \pm \mat{J}_{Y}^{\mu}
  \rightarrow \mat{C}\mat{J}_{\pm}^{\mu}\mat{C}^\dagger.
\end{align*}
In the presence of a finite density, we no longer have manifest
Lorentz invariance and must allow for additional constants into our
spatial derivatives
\begin{equation*}
  \partial^{\mu}_{(v)} = (\partial^0,v\partial^i)
\end{equation*}
to account for the differing speeds of sound.  This paper will be
concerned with static properties, so we can neglect these.  In
principle, one must also match these coefficients $v$.  In QCD this
matching, along with other coefficients, has been made with
perturbative calculations at asymptotic
densities~\cite{Son:1999cm,Son:2000tu}.  Our theory and states still
maintain rotational invariance.  Thus, to lowest-order we
have~\cite{Casalbuoni:1999wu}
\begin{align*}
  \mathcal{L}_{\text{eff}}
  =&\mathcal{L}_{\eta'}+\mathcal{L}_{\Sigma}+\mathcal{L}_{H}+\mathcal{L}_{\phi} + \cdots,\\
  =&\hphantom{+}
  \frac{3f_{\eta'}^2}{4}\partial^{\mu}_{(v_{\eta'})}A^\dagger\partial_{(v_{\eta'})\mu}A
  -\frac{f_{\pi}^2}{4} \tr[\mat{J}_-^{\mu}\mat{J}_{-\mu}]+\\
  &+\frac{3f_{H}^2}{4}\partial_{(v_{H})}^{\mu}V^\dagger\partial_{(v_{H})\mu}V
  -\frac{f_{\phi}^2}{4} \tr[\mat{J}_+^{\mu}\mat{J}_{+\mu}]+\cdots,\\
  =&\tfrac{1}{2}\partial^{\mu}_{(v_{\eta'})}\eta'
  \partial_{(v_{\eta'})\mu} \eta' 
  +\tfrac{1}{2}\partial^{\mu}_{(v_{\pi})} \pi^a 
  \partial_{(v_{\pi})\mu} \pi^a+\\
  &+\tfrac{1}{2}\partial^{\mu}_{(v_{H})} H 
  \partial_{(v_{H})\mu} H
  +\tfrac{1}{2}\partial^{\mu}_{(v_{\phi})} \phi^a
  \partial_{(v_{\phi})\mu} \phi^a
  +\cdots.
\end{align*}
The neglected terms are of higher order in the derivative expansion.
Note that our normalizations have been chosen so that this expression
is canonically normalized to quadratic order in terms of the
dimensionful fields.

The division of $\mathcal{L}_{\text{eff}}$ is
natural~\cite{Casalbuoni:1999wu} because it separates out the colour
singlets.  $\mathcal{L}_{\Sigma}$ depends only on $\mat{\Sigma}$ for
example:
\begin{equation*}
  \mathcal{L}_{\Sigma} = 
  -\frac{f_{\pi}^2}{4}\tr[\mat{J}_-^{\mu}\mat{J}_{-\mu}]
  =
  \frac{f_{\pi}^2}{4}
  \tr[\partial^{\mu}_{(v_{\pi})}\mat{\Sigma}^\dagger\partial_{(v_{\pi})\mu}\mat{\Sigma}].
\end{equation*}
Thus, with the exceptions noted above, the lowest-order low-energy
effective theory of massless $N_f=3$ QCD is
\begin{equation}
  \mathcal{L}_{\text{QCD}} = 
  \mathcal{L}_{\Sigma}+
  \mathcal{L}_{H}+
  \mathcal{L}_{\eta'}+\cdots
\end{equation}
whereas the NJL model proper must also include $\mathcal{L}_{\phi}$.
\subsection{Perturbations}
We shall now consider two types of perturbations: chemical potentials
and quark masses.  To deal with these perturbations, we note that they
enter the microscopic Lagrangian as
\begin{equation*}
  \mathcal{L}_{\text{SB}} = 
  \psi_L^\dagger \mat{\mu}^L \psi_L +
  \psi_R^\dagger \mat{\mu}^R \psi_R +
  \psi_R^\dagger \mat{M} \psi_L +
  \psi_L^\dagger \mat{M}^\dagger \psi_R.
\end{equation*}
These terms break the original symmetries of the theory, but one can
restore these symmetries by imparting the following spurion
transformations to the masses and chemical potentials
\begin{subequations}
  \begin{align}
    \mat{M} &\rightarrow 
    \pm
    (\mat{F}_R^*\otimes\mat{C})\mat{M}(\mat{F}_L^*\otimes\mat{C})^\dagger
    e^{-i(\theta_R-\theta_L)},\\
    \mat{\mu}^L &\rightarrow  
    \hphantom{\pm}(\mat{F}_L^*\otimes\mat{C})\mat{\mu}^L(\mat{F}_L^*\otimes\mat{C})^\dagger,\\
    \mat{\mu}^R &\rightarrow  
    \hphantom{\pm}(\mat{F}_R^*\otimes\mat{C})\mat{\mu}^R(\mat{F}_R^*\otimes\mat{C})^\dagger.
  \end{align}
\end{subequations}
The transformation $\mat{M}\rightarrow -\mat{M}$ preserves the
residual $\text{Z}_2$ symmetries.  This prevents odd powers of the
mass terms from appearing in the chiral effective theory.  In
particular, the linear term dominant in the vacuum is forbidden,
resulting in an inverse mass-ordering of the
mesons~\cite{Son:1999cm,Son:2000tu} with the kaon being the lightest
particle at high density.

All these symmetries must be restored in the effective theory: we are
only allowed to couple these parameters to the fields in ways that
preserve the global symmetries.  To lowest-order, this greatly limits
the possible terms in the effective theory.

In the case of the chemical potentials, we can go one step further by
noting that the perturbations always appears in combination with the
time derivative
\begin{equation}
  \mathcal{L} = \psi^\dagger(i\partial_0+\mat{\mu})\psi+\cdots.
\end{equation}
One can thus promote the chemical potentials to a temporal component
of a spurion gauge field and render the symmetries local in time:
\begin{equation}
  \mat{\mu} \rightarrow (\mat{F}^*\otimes\mat{C})\left(
    \mat{\mu}+i\partial_0
  \right)(\mat{F}^*\otimes\mat{C})^\dagger.
\end{equation}
The effective theory must also maintain these local symmetries.  One
concludes that the chemical potential perturbations can only appear
through the introduction of covariant derivatives in the effective
theory.  In particular, consider adding independent colour and flavour
chemical potential terms:
\begin{equation}
  \mat{\mu}^{L,R} = \mu_{L,R}\mat{1}\otimes\mat{1} 
  + \mat{\mu}^{L,R}_F \otimes \mat{1}
  + \mat{1}\otimes\mat{\mu}_C
\end{equation}
where $\mat{\mu}_F$ and $\mat{\mu}_C$ are traceless $3\times 3$
matrices.  From these we may construct the following quantities that
transform covariantly:
\begin{subequations}
  \label{eq:Nabla0}
  \begin{align}
    \nabla_0\mat{X} &= 
    \partial_0\mat{X}
    +i[\mat{\mu}^L_{F}]^*\mat{X}+i\mat{X}\mat{\mu}_C,\\
    \nabla_0\mat{Y} &= 
    \partial_0\mat{Y}
    +i[\mat{\mu}^{R}_{F}]^*\mat{Y}+i\mat{Y}\mat{\mu}_C,\\
    \nabla_0\mat{\Sigma} &=
    \partial_0\mat{\Sigma}
    +i[\mat{\mu}^{L}_{F}]^*\mat{\Sigma}
    -i\mat{\Sigma}[\mat{\mu}^{R}_{F}]^{T},\\
    \nabla_0 V &= \left(\partial_0 + 2i\mu_{V}\right)V,\\
    \nabla_0 A &= \left(\partial_0 + 2i\mu_{A}\right)A,
  \end{align}
\end{subequations}
where $\mu_{V} = \mu_{R}+\mu_{L}$ is a small adjustment of the
baryon chemical potential $\mu_B/3$ and $\mu_{A} =
\mu_{R}-\mu_{L}$ is the ``axial baryon'' chemical potential.  For
the rest of this paper, we shall only consider vector chemical
potentials that are real and symmetric: $\mat{\mu}^{L,R}_F =
\mat{\mu}_F = \mat{\mu}_F^* = \mat{\mu}_F^\dagger$ etc.  With these
restrictions, the static potential in the effective theory is
\begin{multline}
  \label{eq:Veff}
  V=
  \frac{f_\pi^2}{2}
  \tr[\mat{\Sigma}^\dagger\mat{\mu}_F\mat{\Sigma}\mat{\mu}_F
  -\mat{\mu}_F^2]
  -3f_{H}^2[\mu_V]^2-3f_{\eta'}^2[\mu_A]^2+\\
  +\frac{f_{\phi}^2}{4}
  \tr\left[
    \left(
      \mat{X}^\dagger\mat{\mu}_F\mat{X}
      +\mat{Y}^\dagger\mat{\mu}_F\mat{Y}
      +2\mat{\mu}_C\right)^2
  \right]+\cdots
\end{multline}
to lowest-order.  The terms omitted include terms of higher order in
the perturbation and small corrections due to the explicit violation
of the ``local'' spurion symmetries by the cutoff.
\subsection{Charge Neutrality}
\label{sec:charge-neutrality}
As discussed in~\cite{Alford:2002kj,Kryjevski:2003cu,Gerhold:2003js},
the gauge-invariance of QCD implies that homogeneous states must be
colour neutral.  Non-zero static colour sources
$\mat{A}^{0}_{C}(\vect{p}\!=\!0)$ cancel the tadpole diagrams ensuring
neutrality.  These sources enter the NJL calculation as Lagrange
multipliers to enforce neutrality.

One can see explicitly how these arise in the context of the effective
theory.  The gauge fields effect the local symmetry and thus couple
through the derivatives in exactly the same way as the spurion
coloured chemical potentials: $\mat{\mu}_C \propto
g_s\mat{A}^{0}_{C}$.  Enforcing gauge-invariance induces an effective
coloured chemical potential that makes~(\ref{eq:Veff}) stationary with
respect to variations of the gauge field, and thus equivalently,with
respect to traceless variations of $\mat{\mu}_C$.  Thus, we see that,
to
lowest-order~\cite{Casalbuoni:1999wu,Kryjevski:2003cu,Gerhold:2003js}
\begin{equation}
  \label{eq:muC}
  \mat{\mu}_C = - \tfrac{1}{2}\left(
    \mat{X}^\dagger\mat{\mu}_F\mat{X}+
    \mat{Y}^\dagger\mat{\mu}_F\mat{Y}
  \right).
\end{equation}
Inserting this into the~(\ref{eq:Veff}), and considering only
traceless perturbations, we see that the colour dependence drops out
of the effective theory and we are left with the static effective
potential involving only the colour singlet fields:
\begin{equation}
  V= \frac{f_\pi^2}{2}
  \tr[\mat{\Sigma}^\dagger\mat{\mu}_F\mat{\Sigma}\mat{\mu}_F-\mat{\mu}_F^2]
  +\cdots.
\end{equation}
In order to reproduce the physics of this in the NJL model, however,
we must remove the coloured degrees of freedom.  This is done by
introducing colour chemical potentials to the NJL model as Lagrange
multipliers and using them to impose colour
neutrality~\cite{Alford:2002kj,Kryjevski:2003cu,Gerhold:2003js}.  This
removes the colour dependence in the NJL model to all orders in the
same way as it removes the colour dependence in~(\ref{eq:muC}) to
lowest-order.  (In general, it is not sufficient to impose colour
neutrality: one must also project onto colour singlet states (as well
as states of definite baryon number).  This projection is important
for small systems, but likely has negligible cost for
thermodynamically large systems such as neutron stars.
See~\cite{Amore:2001uf} for an explicit demonstration of this in the
two-flavour case.)

The quarks also couple to the photon, and so we also must enforce
electric neutrality.  Enforcing electromagnetic gauge-invariance will
likewise induce an electric chemical potential $\mu_e$ that ensures
electric neutrality.  It turns out that both the CFL and the
CFLK$^{\text{0}}$ quark matter are neutral under a residual charge
$\tilde{Q}$ (both are $\tilde{Q}$ insulators). This means that one has
some freedom in choosing the chemical potentials used to enforce
neutrality.  In particular, prior to the onset of gapless modes, one
may choose these combinations so that $\mu_e=0$.  This is naturally
enforced by including charged leptons in the calculation.

Once a charged excitation becomes gapless, the material becomes a
conductor and a non-zero $\mu_e$ is required to enforce neutrality.
The phase transition to the gCFL and gCFLK$^\text{0}$ is defined by
exactly such a charged excitation.  In this paper, we shall only
consider the insulating phases, and thus simply set $\mu_e=0$.  For
further discussions of the metal/insulator properties of the CFL and
gCFL we refer the reader
to~\cite{Alford:2003fq,Ruster:2004eg,Fukushima:2004zq}.

\subsection{Kaon Condensation}
\begin{figure}[t]
  {\small
    \psfrag{YYYYYlabelYYYYYY}
    {$(\Omega_{K^0}-\Omega_{CFL})/(\mu_q\mu_{Y}^c)^2$}
    \psfrag{(muY/muYc)2}{$(\mu_Y/\mu_Y^c)^2$}
    \psfrag{0}{$0$}
    \psfrag{-0.01}{$-0.01$}
    \psfrag{-0.02}{$-0.02$}
    \psfrag{-0.03}{$-0.03$}
    \psfrag{-0.04}{$-0.04$}
    \psfrag{0.5}{$0.5$}
    \psfrag{1.0}{$1.0$}
    \psfrag{1.5}{$1.5$}
    \includegraphics{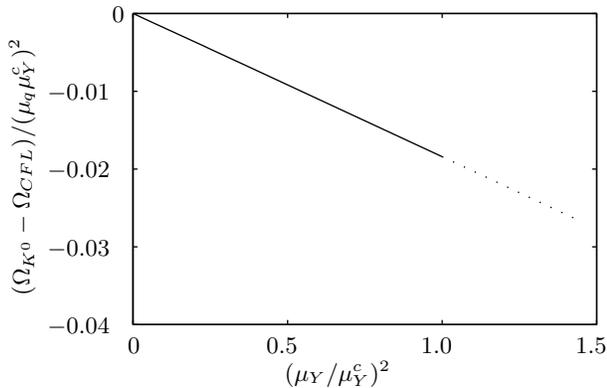}
    \caption{\label{fig:fpiY} Numerical difference in energy densities
      between the kaon-condensed CFLK$^{\text{0}}$ state and the CFL
      state at finite hypercharge potential $\mu_Y$ obtained from our
      microscopic NJL calculation.  The units are scaled in terms of the
      quark chemical potential $\mu_q = 500$ MeV and the critical
      hypercharge chemical potential $\mu_Y^c = 25$ MeV.  The quantities
      plotted were chosen so that the relationship will be linear if our
      calculation agrees with the effective theory
      result~(\ref{eq:fpiMatch}).  The slope of the line is $m =
      -f_{\pi}^2/2\mu_q^2 \approx -0.018$ from which we can determine
      the effective theory parameter $f_\pi \approx 0.19\mu_q$.  This is
      in good numerical agreement with the perturbative QCD result
      $f_\pi \approx 0.209\mu_q$~\cite{Son:1999cm,Son:2000tu}.  The
      dashed extension shows the comparison between the CFLK$^0$
      potential and the CFL potential, but beyond $1.0$, the CFL becomes
      the gCFL and the energy dependence changes.  We have not calculated
      the gCFL potential in this paper, but plot this extension to
      emphasis that the CFLK$^0$ persists beyond the CFL/gCFL transition
      point at $1.0$.}}
\end{figure}
\begin{figure}[t]
  {\small
    \psfrag{YYYYYlabelYYYYYY}
    {$(\Omega_{K^0}-\Omega_{CFL})/(\mu_q\mu_{Y}^c)^2$}
    \psfrag{(muY/muYc)2}{$[M_s^2/(2\mu_q\mu_{Y}^c)]^2$}
    \psfrag{0}{$0$}
    \psfrag{-0.01}{$-0.01$}
    \psfrag{-0.02}{$-0.02$}
    \psfrag{-0.03}{$-0.03$}
    \psfrag{-0.04}{$-0.04$}
    \psfrag{0.5}{$0.5$}
    \psfrag{1.0}{$1.0$}
    \psfrag{1.5}{$1.5$}
    \includegraphics{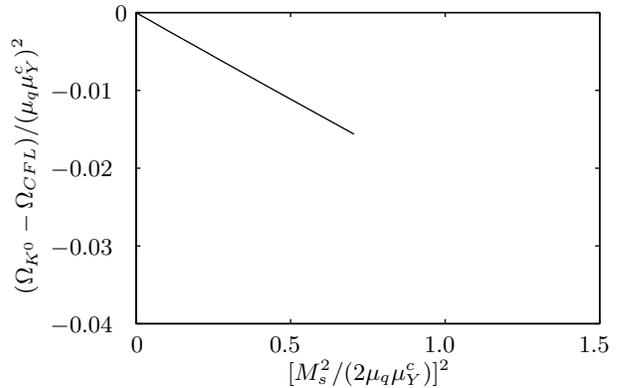}
    \caption{\label{fig:fpims} Numerical difference in energy densities
      between the kaon-condensed CFLK$^{\text{0}}$ state and the CFL
      state at finite strange quark mass $m_s$ obtained from our
      microscopic NJL calculation.  The units are scaled in terms of the
      renormalized quark chemical potential $\mu_q \approx 500$ MeV and
      the critical hypercharge chemical potential $\mu_Y^c = 25$ MeV to
      facilitate comparison with Figure~\ref{fig:fpiY} and to emphasize
      the linear relationship implicit in~(\ref{eq:fpimsMatch1}).  The
      slope of the line is $m \approx -f_{\pi}^2/2\mu_q^2 \approx
      -0.028$ which gives an effective $f_\pi \approx 0.21\mu_q$ which
      is consistent with our previous results.  In comparison with
      Figure~\ref{fig:fpiY}, the CFL$\rightarrow$gCFL transition occurs
      somewhat earlier because the gap parameters are reduced with
      increasing strange quark mass.  The curve cannot be extended as in
      Figure~\ref{fig:fpiY} because the free-energy of the CFL is no
      longer a constant as it was with a hypercharge perturbation.}}
\end{figure}
We are now in a position to argue for the existence of a
kaon-condensed state.  Consider performing an axial $\text{K}^0$
rotation on the parity-even CFL state.  This is effected
using~(\ref{eq:PsiTransform}) in the microscopic theory and
using~(\ref{eq:EffTransform}) in the effective theory with the
parameter $\tilde{\pi}^6 = \theta$.  Such a state is now described by
\begin{equation}
  \mat{\Sigma} = e^{2 i \theta\mat{\lambda}_6} = \begin{pmatrix}
    1\\
    & \cos(\theta)  & i\sin(\theta)\\
    & i\sin(\theta) & \cos(\theta)
  \end{pmatrix}.
\end{equation}
In the presence of a hypercharge chemical potential, the effective
potential becomes~\cite{Schafer:2000ew,Bedaque:2001je,Kaplan:2001qk}
\begin{equation}
  \label{eq:Vtheta}
  V(\theta) = \frac{f_{\pi}^2 \mu_Y^2}{2}\left(
    \cos^2(\theta)-1
  \right)+\cdots.
\end{equation}
We see that this has a minimum for $\theta=\pm\pi/2$: this is the
state with maximal $\text{K}^0$ condensation.  We can also directly
compute the difference in the thermodynamic potential densities
between the CFL state and the CFLK$^{\text{0}}$ state:
\begin{equation}
  \label{eq:fpiMatch}
  \Omega_{\text{CFLK}^0}-\Omega_{\text{CFL}} = -\frac{f_{\pi}^2 \mu_Y^2}{2}.
\end{equation}
Armed with this relationship, we can now turn to the microscopic
calculation and determine the coefficient $f_{\pi}$.  In
Figure~\ref{fig:fpiY} we plot our numerical results so that the
linear relationship~(\ref{eq:fpiMatch}) is evident.  From the slope of
the relationship we find that
\begin{equation}
  f_{\pi} \approx 0.19\mu_q.
\end{equation}
We note that this is in good numerical agreement with the perturbative
QCD result~\cite{Son:1999cm,Son:2000tu} of $f_{\pi} = 0.209\mu_q$.
This striking agreement between two very different models arises from
the fact that this coefficient is not very sensitive to the effects of
the cutoff (which is different in the two theories) and gives
encouraging support to the use of the NJL model to study QCD.

The equivalent relationship in the case of a strange quark mass
requires one to include mass terms in the effective theory (see for
example~\cite{Schafer:2001za}), but the leading order effect can be
determined by using the ``effective'' strange quark chemical potential
$\mu_Y \approx M_s^2/(2\mu_q)$ that follows from~(\ref{eq:ms=muy1}):
\begin{equation}
  \label{eq:fpimsMatch1}
  \Omega_{\text{CFLK}^0}-\Omega_{\text{CFL}} = 
  -\frac{f_{\pi}^2}{2}\left(\frac{M_s^2}{2\mu_q}\right)^2.
\end{equation}
It is important here to note, however, that the strange quark mass
affects the solution in such a way that the gap parameters change and
self-energy corrections modify the quark chemical potential and the
constituent quark mass.  It is the renormalized parameters that appear
in this relation and in the perturbative QCD result.  Thus, as a
function of the bare parameter $m_s$, we have $M_s \propto m_s$ and
$\mu_q - \mu_s \propto m_s^2$.  Thus, we should see a linear
relationship between $\Omega_{\text{CFLK}^0}-\Omega_{\text{CFL}}$ and
$M_s^4$.  We plot this relationship in Figure~\ref{fig:fpims} and
extract the slope which gives the relationship $f_\pi \approx 0.21
\mu_q$.  This is in qualitative agreement with our previous result.
The slight numerical disagreement is due to effects of the strange
quark mass that are not captured by the chemical potential
shift~(\ref{eq:ms=muy1}).
\begin{figure}[t]
  {\small
    \psfrag{YYYYYlabelYYYYYY}
    {$-2 \mu_3/\mu_Y = -4\mu_8/\mu_Y$}
    \psfrag{muY/muYc}{$\mu_Y/\mu_Y^c$}
    \psfrag{0}{$0$}
    \psfrag{0.5}{$0.5$}
    \psfrag{2.0}{$2.0$}
    \psfrag{0.8}{$0.8$}
    \psfrag{1.0}{$1.0$}
    \psfrag{1.2}{$1.2$}
    \psfrag{1.5}{$1.5$}
    \includegraphics{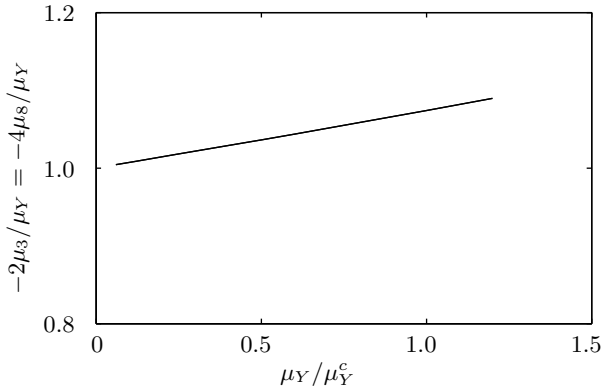}
    \caption{\label{fig:muColour} Chemical potentials required by the
      NJL model to enforce colour neutrality in the CFLK$^{0}$ phase
      with finite hypercharge chemical potentials.  Note that the
      effective theory relationship~(\ref{eq:muColour}) is satisfied
      from small chemical potentials.  The linear deviation seen here
      reflects the missing terms in the effective theory that are of
      higher order in the perturbation $\mu_Y$ with a linear deviation
      here corresponds to $\mu_Y^2$ terms missing
      in~(\ref{eq:muColour}).}}
\end{figure}

We pause here to point out a discrepancy between our results and
similar work by Buballa~\cite{Buballa:2004sx}.  Our results shown in
Figure~\ref{fig:fpims} suggests that kaon condensation occurs for all
values of $m_s$ in this simple model with $m_u=m_d=0$ whereas Buballa
finds that kaon condensation is only favoured for $m_s$ sufficiently
large.  If the chemical potential shift were the only effect of a
strange quark mass, then this would be inconsistent
with~(\ref{eq:Veff}).  This is not, however, the correct expansion.
Instead, one has, for maximal kaon condensation and $m_u=m_d=0$:
\begin{equation}
  \label{eq:MassTerms}
  \Omega_{\text{CFLK}^0}-\Omega_{\text{CFL}} = 
  -(4a_6+a_8)m_s^2 -c m_s^4 + \order(m_s^6).
\end{equation}
A proper discussion of this is beyond the scope of this paper, but
will be discussed elsewhere~\cite{FK:2005}.  The term with coefficient
$c$ should be identified with the leading order contribution
from~(\ref{eq:ms=muy1}); the term with coefficient $a_6$ arises from
the sextet gap contribution $\Delta_6$ and is discussed
in~\cite{Kryjevski:2004cw}; the term with coefficient $a_8$ is of
higher order in perturbative QCD and so is usually
neglected~\cite{Son:1999cm,Son:2000tu,Schafer:2001za}.

Thus, Buballa's results are consistent with the effective theory.  The
discrepancy is due to a different choice of parameters $\Delta \sim
100$ MeV and $\Lambda = 600$ MeV compared with our parameters $\Delta
\sim 25$ MeV and $\Lambda = 800$ MeV.  With the large gap, one is
further from the perturbative QCD regime and the quadratic term
appears to play a significant role.  In our analysis, the term
$-(4a_6+a_8)m_s^2$ is small compared with the $c m_s^4$ term.  Using
Buballa's parameters, however, we qualitatively reproduce his results.
A further discussion of these effects will be presented
shortly~\cite{FK:2005}.
\begin{figure}[t]
  {\small
    \psfrag{YYlabelYYY}
    {$n_Y/(\mu_Y^c\mu_q^2)$}
    \psfrag{muY/muYc}{$\mu_Y/\mu_Y^c$}
    \psfrag{0}{$0$}
    \psfrag{0.05}{$0.05$}
    \psfrag{0.5}{$0.5$}
    \psfrag{1.0}{$1.0$}
    \psfrag{1.5}{$1.5$}
    \includegraphics{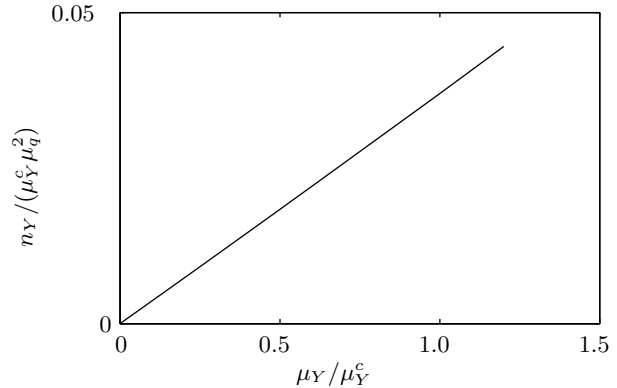}
    \caption{\label{fig:nY} Hypercharge density of the CFLK$^{0}$ state
      in the presence of a hypercharge chemical potential $\mu_Y$ as
      obtained from our microscopic NJL model calculation.  The units
      are scaled as in Figure~\ref{fig:fpiY} so that the relationship
      will be linear if the NJL model calculation agrees with the
      effective theory prediction~(\ref{eq:nY}).  By determining the
      slope of this relationship we have another way of determining the
      coefficient $f_{\pi}$ in the effective theory.  The slope is
      $f_{\pi}^2/\mu_q^2 \approx 0.037$ which agrees with our previous
      determination of $f_\pi \approx 0.19\mu_q$.}}
\end{figure}

There are a couple of other consequences that follow directly from the
effective theory.  One is the value of the coloured chemical
potentials required to enforce neutrality.  In our microscopic model,
we have fixed the gauge (unitary gauge) by setting $\mat{X} =
\mat{Y}^\dagger = \sqrt{\mat{\Sigma}}$ for the axial rotations.  The
CFL state has $\mat{X}=\mat{Y}=\mat{1}$ while the CFLK$^{\text{0}}$
state has
\begin{equation}
  \mat{X} = \mat{Y}^\dagger = 
  \frac{1}{\sqrt{2}}
  \begin{pmatrix}
    \sqrt{2}\\
    &1 & i\\
    & i & 1
  \end{pmatrix}.
\end{equation}
From~(\ref{eq:muC}) we have the following relationships required to
enforce neutrality~\cite{Kryjevski:2003cu}
\begin{subequations}
  \label{eq:muColour}
  \begin{align}
    \mu_8 &= -\mu_Y, & \mu_3 &= 0, & &\text{(CFL)},\\
    \mu_8 &= -\frac{1}{4}\mu_Y, & \mu_3 &= -\frac{1}{2}\mu_Y, &
    &\text{(CFLK$^{\text{0}}$)}.
  \end{align}
\end{subequations}
We plot these relationships in Figure~\ref{fig:muColour}.  Note that
they only hold for small perturbations where the effective theory is
valid: this plot also demonstrates a departure from the lowest-order
effective theory as the perturbation is increased.

As a final demonstration of the effective theory, we calculate the
hypercharge density.  This is obtained by varying the thermodynamic
potential with respect to the hypercharge chemical potential:
\begin{equation}
  \label{eq:nY}
  n_Y = -\pdiff{\Omega}{\mu_Y} \approx -f_\pi^2\mu_Y(\cos^2(\theta)-1).
\end{equation}
There should be no hypercharge density in the CFL state and a density
of $n_Y = f_\pi^2\mu_Y$ in the CFLK$^0$ state.  Indeed, the CFL
supports no hypercharge density with $n_u = n_d = n_s$.  The
hypercharge density of the CFLK$^{0}$ phase is shown in
Figure~\ref{fig:nY} and provides another method of extracting $f_{\pi}
= 0.19\mu_q$.
\subsection{A Note on the Meaning of $V(\theta)$}
We make a few remarks here about the meaning of the effective
potential $V(\theta)$.  In particular, one might be tempted to try and
compute the functional form of $V(\theta)$ in the microscopic theory
to facilitate matching with the effective theory.  Such an approach
will generally fail because one is allowed to pick an arbitrary
parametrization of the Goldstone fields as long as they leave the
kinetic terms unaltered~\cite{Coleman:1969sm,Callan:1969sn}.  Physical
quantities must be invariant under this change of parametrization:
thus the spectrum about the minimum, densities, and energy differences
are reasonable quantities to compare in each theory.  The general form
of the effective potential away from the stationary points, however,
is rather arbitrary.

As an example: consider starting with the parity even CFL state in the
presence of a finite $\mu_Y$.  This state corresponds to a stationary
point of the effective potential and is a self-consistent solution to
the gap equations.  One can then form a continuum of ``kaon rotated''
states $\ket{\theta}$ by applying the broken symmetry generators to
this state.  One might expect to find $V(\theta)$ by computing the
energy of these states, but instead one finds an expression that is
only valid locally about the stationary point.  The reason is twofold:
First, there is not a unique ``kaon rotated'' state $\ket{\theta}$.
This state has many other parameters corresponding to other
``directions'' (such as the gap parameters $\Delta$, the chemical
potential corrections etc.)  The only way to uniquely determine these
is to solve the gap equations, and these only have well-defined
solutions at stationary points. Second, the generators of the
pseudo-Goldstone bosons in the presence of perturbations are not the
same as the generators of the true Goldstone bosons in the unbroken
model: the pseudo-Goldstone bosons have some admixture of these other
``directions''.

This becomes even more evident when you analyze the CFLK$^0$ state
with a large perturbation: one can try to ``undo'' the kaon rotation
by applying the appropriate symmetry generators to minimize the parity
violating condensates, but one finds that there is no way to do this.
One must also transform the other parameters in order convert a
CFLK$^0$ state back to a parity even CFL state.

\section{Conclusion}
We have explicitly found self-consistent solutions within a
microscopic NJL model exhibiting the feature of kaon condensation in a
colour-flavour--locked state.  Using these solutions, we have
demonstrated that by properly enforcing gauge neutrality, one can
remove the extraneous coloured degrees of freedom from the NJL model
and effectively model kaon condensation in high-density QCD.  In
particular, the microscopic calculations can be matched onto the
low-energy effective theory of QCD.  We determined $f_\pi = 0.19\mu_q$
which is in good numerical agreement with the perturbative QCD result.

Furthermore, our solutions are fully self-consistent: no
approximations have been made beyond the mean-field approximation and
restricting our attention to isotropic and homogeneous states.  We
find that our results agree qualitatively with both the expected
properties of the CFLK$^0$ phase based on effective theory
calculations, and with the previous numerical calculations of the
CFL/gCFL transition.

Quantitatively we find that the phase transitions occur at slightly
smaller parameter values than previously found in the literature.
Concerning the CFL/gCFL transition, we find that the gap parameters
are reduced by a few percent compared with those presented
in~\cite{Fukushima:2004zq}, and subsequently, the critical $M_s$ is
also a few percent lower.  Concerning the CFLK$^0$/gCFLK$^0$
transition, we find that the transition occurs about a factor of
$1.2$ higher than the CFL/gCFL transition.  This is in qualitative
agreement but quantitative disagreement with the factor of $4/3$
calculated in~\cite{Kryjevski:2004kt}.

The next step is to use this microscopic model to determine the phase
structure of high-density QCD in the region where the gapless modes
appear.  We suspect that the gCFLK$^0$ state will survive somewhat
longer than the gCFL state on account of its lower condensation
energy, but a quantitative comparison is required.  Extrapolation to
finite temperature is also a trivial extension in our formalism.

A somewhat more challenging direction is to consider the effects of
instantons and finite up and down quark masses and investigate other
forms of meson condensation.  Preliminary investigations indicate,
however, that the number of parameters required to close the gap
equations in the presence of arbitrary meson rotations may be
prohibitively large to continue with fully self-consistent
calculations.  This should still be tractable with carefully made
approximations.
\section{Acknowledgements}
I would like to thank K.~Rajagopal and F.~Wilczek for suggesting this
problem, and P.~Bedaque, K.~Fukushima, D.~K.\ Hong, D.~Kaplan,
C.~Kouvaris, J.~Kundu, M.~J.\ Savage, T.~Sch\"{a}fer, I.~Shovkovy, I.~Stewart, and
A.~R.\ Zhitnitsky for useful discussions.  Related work has been done
independently by M.~Buballa~\cite{Buballa:2004sx} and I am grateful to
him for postponing his preprint submission while the first version of
this work was completed.  This work is supported in part by funds
provided by the U.S.\ Department of Energy (D.O.E.) under cooperative
research agreement \#DF-FC02-94ER40818.

%\bibliographystyle{h-physrev3}
%\bibliography{master}

\newpage
\appendix
\section{Full Parametrization}
\label{sec:full-param}
In this appendix, we give the full parametrization used to analyze
the K$^0$ condensed states.  First, we must introduce a full set of
diagonal chemical potentials.  One approach would be to introduce the
$9$ individual quark chemical potentials, but certain linear
combinations couple to relevant physics.  We fix the overall density
by fixing the baryon chemical potential $\mu_B$.  Then we must enforce
gauge neutrality, so we introduce $\mu_e$ which couples to the
electromagnetic field, and the diagonal colour chemical potentials
$\mu_3$ and $\mu_8$.  The rest of the chemical potentials are chosen
to be orthogonal to these.  Here then are the diagonal elements of the
diagonal chemical potentials expressed as tensor products of the
flavour and colour structure :
\begin{subequations}
  \begin{align}
    \mat{\mu}_B &\times [1,\hphantom{-}1,\hphantom{-}1] \otimes 
    [1,\hphantom{-}1,\hphantom{-}1]/3,\\
    \mat{\mu}_e &\times [2,-1,-1] \otimes
    [1,\hphantom{-}1,\hphantom{-}1]/3,\\
    \mat{\mu}_3 &\times [1,\hphantom{-}1,\hphantom{-}1] \otimes 
    [1,-1,\hphantom{-}0]/2,\\
    \mat{\mu}_8 &\times [1,\hphantom{-}1,\hphantom{-}1] \otimes
    [1,\hphantom{-}1,-2]/3,\\
    \mat{\mu}_{f}&\times [0,\hphantom{-}1,-1] \otimes 
    [1,\hphantom{-}1,\hphantom{-}1],\\
    \mat{\mu}_{e3}&\times [2,-1,-1]\otimes
    [1,-1,\hphantom{-}0],\\
    \mat{\mu}_{e8}&\times [2,-1,-1]\otimes
    [1,\hphantom{-}1,-2],\\
    \mat{\mu}_{f3}&\times [0,\hphantom{-}1,-1]\otimes
    [1,-1,\hphantom{-}0],\\
    \mat{\mu}_{f8}&\times [0,\hphantom{-}1,-1]\otimes
    [1,\hphantom{-}1,-2].
  \end{align}
\end{subequations}
An alternative set of chemical potentials includes the hypercharge
chemical potential $\mu_Y$ instead of $\mu_{f}$.  These are no longer
orthogonal, but are still linearly independent.
\begin{subequations}
  \begin{align}
    \mat{\mu}_B &\times [1,\hphantom{-}1,\hphantom{-}1] \otimes
    [1,\hphantom{-}1,\hphantom{-}1]/3,\\
    \mat{\mu}_e &\times [2,-1,-1] \otimes
    [1,\hphantom{-}1,\hphantom{-}1]/3,\\
    \mat{\mu}_3 &\times [1,\hphantom{-}1,\hphantom{-}1] \otimes
    [1,-1,\hphantom{-}0]/2,\\
    \mat{\mu}_8 &\times [1,\hphantom{-}1,\hphantom{-}1] \otimes
    [1,\hphantom{-}1,-2]/3,\\
    \mat{\mu}_Y &\times [1,\hphantom{-}1,-2]\otimes
    [1,\hphantom{-}1,\hphantom{-}1]/3,\\
    \mat{\mu}_{e3}&\times [2,-1,-1]\otimes
    [1,-1,\hphantom{-}0],\\
    \mat{\mu}_{e8}&\times [2,-1,-1]\otimes
    [1,\hphantom{-}1,-2],\\
    \mat{\mu}_{f3}&\times [0,\hphantom{-}1,-1]\otimes
    [1,-1,\hphantom{-}0],\\
    \mat{\mu}_{f8}&\times [0,\hphantom{-}1,-1]\otimes
    [1,\hphantom{-}1,-2].
  \end{align}
\end{subequations}
The diagonal mass corrections (chiral condensates) do not couple to
any external physics, so we simply use the nine quark mass corrections
($\delta{}m_{ur}$, $\delta{}m_{ug}$, $\delta{}m_{ub}$,
$\delta{}m_{dr}$, $\delta{}m_{dg}$, $\delta{}m_{db}$,
$\delta{}m_{sr}$, $\delta{}m_{sg}$, $\delta{}m_{sb}$).  

The rest of the parameters are described in the following matrices.
These appear more condensed when expressed in the basis described
in~\cite{Fukushima:2004zq} where the quarks are ordered (ru, gd, bs,
rd, gu, rs, bu, gs, bd).  In this basis, the matrices corresponding to
the variational parameters
\begin{subequations}
  \label{eq:ABdef1}
  \begin{align*}
    \mat{A} &= \mat{1}\otimes \delta\mat{\mu} 
    + \mat{\gamma}_5 \otimes \delta\mat{\mu}_5
    - \mat{\gamma}_0 \otimes \delta\mat{m}
    - \mat{\gamma}_0\mat{\gamma}_5\otimes \delta\mat{m}_5,\\
    \mat{B} &= \mat{\gamma}_C\mat{\gamma}_5\otimes\mat{\Delta} 
    + \mat{\gamma}_C\otimes \mat{\Delta}_5
    + \mat{\gamma}_0\mat{\gamma}_C\mat{\gamma}_5 \otimes \mat{\kappa}
    + \mat{\gamma}_0\mat{\gamma}_C \otimes\mat{\kappa}_5.
  \end{align*}
\end{subequations}
In order to allow for a computer to enumerate the parameters, we
introduce a systematic method for labelling the parameters.  First, we
use one of the names $\mu$, $\mu^5$, $m$, $m^5$, $\Delta$, $\Delta^5$,
$\kappa$, or $\kappa^5$ corresponding to the structure given
above.  We then use a two-digit index to specify which
elements are non-zero and an $i$ indicates that the specified element
is $i$ rather than simply $1$.  The symmetric entry must also be set
so that the resulting matrix $\mat{\gamma}\otimes\mat{\mu}$ is either
Hermitian or anti-symmetric depending on whether or not it
parametrizes $\mat{A}$ or $\mat{B}$ respectively.  In total, there are
$666$ independent matrices.  For example
\begin{subequations}
  \begin{gather*}
    \mat{\mu}_{12}
    =\mat{\mu}^5_{12}
    =\mat{m}_{12}
    =\mat{\Delta}_{12}
    =\mat{\Delta}^5_{12}
    =\mat{\kappa}_{12}
    =
    \begin{pmatrix}
      0&1&0&\cdots\\
      1&0&0\\
      0&0&0\\
      \vdots&&&\ddots
    \end{pmatrix},\\
    \mat{m}^{5}_{12} 
    =\mat{\kappa}^{5}_{12}
    =
    \begin{pmatrix}
      0&1&0&\cdots\\
      -1&0&0\\
      0&0&0\\
      \vdots&&&\ddots
    \end{pmatrix},\\
    \mat{\mu}_{12i}
    =\mat{\mu}^5_{12i}
    =\mat{m}_{12i}
    =\mat{\kappa}^{5}_{12i}
    =
    \begin{pmatrix}
      0&i&0&\cdots\\
      -i&0&0\\
      0&0&0\\
      \vdots&&&\ddots
    \end{pmatrix},\\
    \mat{m}^{5}_{12i} 
    =\mat{\Delta}_{12i}
    =\mat{\Delta}^5_{12i}
    =\mat{\kappa}_{12i}
    =
    \begin{pmatrix}
      0&i&0&\cdots\\
      i&0&0\\
      0&0&0\\
      \vdots&&&\ddots
    \end{pmatrix}.
  \end{gather*}
\end{subequations}
The reason that $\mat{m}^{5}$ and $\mat{\kappa}^5$ behave differently
than the others is that, while $\mat{1}$, $\mat{\gamma}_5$, and
$\mat{\gamma}_0$ are Hermitian, $\mat{\gamma}_0\mat{\gamma}_5$ is
anti-Hermitian.  Likewise, while $\mat{\gamma}_C\mat{\gamma}_5$,
$\mat{\gamma}_C$, and $\mat{\gamma}_0\mat{\gamma}_C\mat{\gamma}_5$ are
anti-symmetric, $\mat{\gamma}_0\mat{\gamma}_C$ is symmetric.  Again,
recall that these are all specified in the ``unlocking'' basis which
is ordered as
\begin{equation}
  \text{ru, gd, bs, rd, gu, rs, bu, gs, bd}.
\end{equation}
The parity even CFL state with no mass or hypercharge is expressed
in terms of this parametrization as 
\begin{subequations}
  \begin{align}
    \Delta_{12} = \Delta_{13} = \Delta_{23} &= (\Delta_3+\Delta_6)/2,\\
    \Delta_{45} = \Delta_{67} = \Delta_{89} &= (\Delta_6-\Delta_3)/2,\\
    \Delta_{11} = \Delta_{22} = \Delta_{33} &= \Delta_6,\\
    \mu_{12} = \mu_{13} = \mu_{23} &=-3\mu_{\text{oct}},\\
    \mu_{e3} = 3\mu_{e8} = -\mu_{f3} = \mu_{f8} &=
    -3\mu_{\text{oct}}/4.
  \end{align}
\end{subequations}
Here are some comparisons with other conventions in the literature.
Alford, Kouvaris, and Rajagopal~\cite{Alford:2003fq} introduce
$\Delta_{1}$, $\Delta_{2}$ and $\Delta_{3}$ which are all related to
the attractive anti-symmetric $\bar{3}$ channel:
\begin{subequations}
  \begin{align}
    \Delta_{23} = \Delta_{89} &= \Delta_{1},\\ 
    \Delta_{13} = \Delta_{67} &= \Delta_{2},\\ 
    \Delta_{12} = \Delta_{45} &= \Delta_{3}. 
  \end{align}
\end{subequations}
R\"uster, Shovkovy, and Rischke~\cite{Ruster:2004eg} introduces the
parameters $\phi$, $\varphi$ and $\sigma$ which include the repulsive
symmetric $6$ channel parameters:
\begin{subequations}
  \begin{align}
    \Delta_{23} &= \varphi_{1}, &
    \Delta_{13} &= \varphi_{2}, &
    \Delta_{12} &= \varphi_{3}, \\
    \Delta_{89} &= \phi_{1}, &
    \Delta_{67} &= \phi_{2}, &
    \Delta_{45} &= \phi_{3},\\
    \Delta_{11} &= \sigma_{1}, &
    \Delta_{22} &= \sigma_{2}, &
    \Delta_{33} &= \sigma_{3}.
  \end{align}
\end{subequations}
Finally, Buballa~\cite{Buballa:2004sx} uses only the following parameters to
parametrize the meson condensed phases:
\begin{subequations}
  \begin{align}
    -\Delta_{12} = \Delta_{45} &= s_{22}/4,\\
    -\Delta_{13} = \Delta_{67} &= s_{55}/4,\\
    -\Delta_{23} = \Delta_{89} &= s_{77}/4,\\
    -\Delta^5_{19i} = \Delta^5_{47i} &= p_{25}/4,\\
    -\Delta^5_{18i} = \Delta^5_{56i} &= p_{52}/4.
  \end{align}
\end{subequations}
In Tables~\ref{tab:CFLY}, \ref{tab:CFLYK0}, \ref{tab:CFLMs}, and
\ref{tab:CFLK0Ms} we give the numerical values of the parameters for
each of the states displayed in Figures~\ref{fig:YDispersions},
\ref{fig:K0YDispersions}, \ref{fig:MSDispersions},
and~\ref{fig:MSK0Dispersions} respectively.  We only list the non-zero
parameters: the other parameters are zero.
\begin{table}[ht]
  \begin{center}
   {\small
     \begin{tabular}{||r||l|l||l|l||}
       \hline
       & \multicolumn{2}{c||}{$\mu_Y = 0.50\mu_Y^c$} 
       & \multicolumn{2}{c||}{$\mu_Y = \mu_Y^c$}\\
       \cline{2-5}
       Param. & Bare & Correction & Bare & Correction\\
       \hline
       $\mu_B/3$ & $+549.93$ & $-49.93$ & $+549.93$ & $-49.93$ \\
       $\mu_8$ & $-12.5$ & $+0$ & $-25$ & $+0$ \\
       $\mu_Y$ & $+12.5$ & $+0$ & $+25$ & $+0$ \\
       $\mu_{oct}$ & $0$ & $-0.031332$ & $0$ & $-0.031332$ \\
       \hline
       $\Delta_{3}$ & $0$ & $+25.657$ & $+$ & $+25.657$ \\
       $\Delta_{6}$ & $0$ & $+0.65709$ & $0$ & $+0.65709$ \\
       \hline
     \end{tabular}
     \caption[Self-consistent parameters for a CFL phase with finite
     $\mu_Y$]{\label{tab:CFLY} Parameters required for a
       self-consistent parity-even CFL solution in the presence of a
       hypercharge chemical potential.  These values correspond to the
       dispersions shown in Figure~\ref{fig:YDispersions}.  All values
       are in MeV.  The first column labelled ``Bare'' gives the fixed
       bare parameters that enter the Hamiltonian~(\ref{eq:H}).  The
       column labelled ``Correction'' is the contribution from the
       self-energy.  The sum of the columns is the value that enters
       the quadratic Hamiltonian~(\ref{eq:H0}).}
   }
 \end{center}
\end{table}
\begin{table*}[ht]
  \begin{center}
    {\small
      \begin{tabular}{||r||l|l||l|l||}
        \hline
        & \multicolumn{2}{c||}{$\mu_Y = 0.50\mu_Y^c$} 
        & \multicolumn{2}{c||}{$\mu_Y = 1.20\mu_Y^c$}\\
        \cline{2-5}
        Param. & Bare & Correction & Bare & Correction\\
        \hline
        $\mu_B/3$ & $+549.93$ & $-49.932$ & $+549.93$ & $-49.947$ \\
        $\mu_Y$ & $+12.5$ & $-1.0687$ & $+30$ & $-2.5931$ \\
        $\mu_e$ & $0$ & $+0.53436$ & $0$ & $+1.2965$ \\
        $\mu_3$ & $-6.4772$ & $-0.00000$ & $-16.346$ & $-3.331\times 10^{-7}$ \\
        $\mu_8$ & $-3.2386$ & $-0.00000$ & $-8.1729$ & $-1.665\times 10^{-7}$ \\
        $\mu_{e3}$ & $0$ & $+0.02421$ & $0$ & $+0.027856$ \\
        $\mu_{e8}$ & $0$ & $+0.0080699$ & $0$ & $+0.0092853$ \\
        $\mu_{f3}$ & $0$ & $+0.035998$ & $0$ & $+0.084801$ \\
        $\mu_{f8}$ & $0$ & $+0.011999$ & $0$ & $+0.028267$ \\
        $\mu_{12}=\mu^5_{18i}$ & $0$ & $+0.016852$ & $0$ & $-0.047357$ \\
        $\mu_{13}=\mu^5_{19i}$ & $0$ & $+0.11902$ & $0$ & $+0.19694$ \\
        $\mu_{23}=\mu_{89}$ & $0$ & $+0.046967$ & $0$ & $+0.046617$ \\
        $\mu^5_{28i}$ & $0$ & $+0.088616$ & $0$ & $+0.14526$ \\
        $\mu^5_{29i}=\mu^5_{38i}$ & $0$ & $+0.046967$ & $0$ & $+0.046617$ \\
        $\mu^5_{39i}$ & $0$ & $+0.0030234$ & $0$ & $-0.065811$ \\
        $\mu^5_{46i}$ & $0$ & $+0.11479$ & $0$ & $+0.27514$ \\
        \hline
        $\Delta_{11}$ & $0$ & $+0.64468$ & $0$ & $+0.5851$ \\
        $\Delta_{22}=-\Delta_{88}=-\Delta^5_{28i}$ & $0$ & $+0.32265$ & $0$ & $+0.31566$ \\
        $\Delta_{33}=-\Delta_{99}=-\Delta^5_{39i}$ & $0$ & $+0.33523$ & $0$ & $+0.34523$ \\
        $\Delta_{12}=-\Delta^5_{18i}$ & $0$ & $+9.6383$ & $0$ & $+10.138$ \\
        $\Delta_{13}=-\Delta^5_{19i}$ & $0$ & $+8.9893$ & $0$ & $+8.5746$ \\
        $\Delta_{23}=-\Delta_{89}$ & $0$ & $+12.789$ & $0$ & $+12.605$ \\
        $\Delta_{45}=-\Delta^5_{56i}$ & $0$ & $-9.1811$ & $0$ & $-9.7012$ \\
        $\Delta_{67}=-\Delta^5_{47i}$ & $0$ & $-8.5229$ & $0$ & $-8.1128$ \\
        $\Delta^5_{29i}=\Delta^5_{38i}$ & $0$ & $-0.33242$ & $0$ & $-0.35025$ \\
        \hline
      \end{tabular}
      \caption[Self-consistent parameters for a CFLK$0$ phase with
      finite $\mu_Y$]{\label{tab:CFLYK0} Parameters required for a
        self-consistent CFLK$0$ solution in the presence of a
        hypercharge chemical potential.  These values correspond to the
        dispersions shown in Figure~\ref{fig:K0YDispersions}.  All
        values are in MeV.  The first column labelled ``Bare'' gives the
        fixed bare parameters that enter the Hamiltonian~(\ref{eq:H}).
        The column labelled ``Correction'' is the contribution from the
        self-energy.  The sum of the columns is the value that enters
        the quadratic Hamiltonian~(\ref{eq:H0}).}
    }
  \end{center}
\end{table*}
\begin{table*}[ht]
  \begin{center}
    {\small
      \begin{tabular}{||r||l|l||l|l||}
        \hline
        & \multicolumn{2}{c||}{$M_s^2/(2\mu) = 0.50\mu_Y^c$} 
        & \multicolumn{2}{c||}{$M_s^2/(2\mu) = 0.83\mu_Y^c$}\\
        \cline{2-5}
        Param. & Bare & Correction & Bare & Correction\\
        \hline 
        $\mu_B/3$ & $+549.93$ & $-48.952$ & $+549.93$ & $-48.316$ \\
        $\mu_8$ & $-12.649$ & $-0.00000$ & $-20.95$ & $+1.469\times 10^{-7}$ \\
        $\mu_{e3}$ & $0$ & $+0.022617$ & $0$ & $+0.022026$ \\
        $\mu_{e8}$ & $0$ & $+0.0074542$ & $0$ & $+0.0072087$ \\
        $\mu_{f3}$ & $0$ & $-0.022617$ & $0$ & $-0.022026$ \\
        $\mu_{f8}$ & $0$ & $+0.022362$ & $0$ & $+0.021626$ \\
        $\mu_{12}$ & $0$ & $+0.090467$ & $0$ & $+0.088103$ \\
        $\mu_{13}=\mu_{23}$ & $0$ & $+0.088963$ & $0$ & $+0.085669$ \\
        $m_{ur}=m_{dg}$ & $0$ & $+0.15778$ & $0$ & $+0.19366$ \\
        $m_{ug}=m_{dr}$ & $0$ & $+0.17255$ & $0$ & $+0.2117$ \\
        $m_{ub}=m_{db}$ & $0$ & $+0.15604$ & $0$ & $+0.19155$ \\
        $m_{sr}=m_{sg}$ & $+61.843$ & $+50.029$ & $+80$ & $+64.267$ \\
        $m_{sb}$ & $+61.843$ & $+50.079$ & $+80$ & $+64.329$ \\
        $m_{12}$ & $0$ & $-0.014765$ & $0$ & $-0.018037$ \\
        $m_{13}=m_{23}$ & $0$ & $+0.026496$ & $0$ & $+0.032723$ \\
        \hline
        $\Delta_{11}=\Delta_{22}$ & $0$ & $-0.62077$ & $0$ & $-0.59745$ \\
        $\Delta_{33}$ & $0$ & $-0.64043$ & $0$ & $-0.62926$ \\
        $\Delta_{12}$ & $0$ & $-12.914$ & $0$ & $-12.753$ \\
        $\Delta_{13}=\Delta_{23}$ & $0$ & $-12.639$ & $0$ & $-12.302$ \\
        $\Delta_{45}$ & $0$ & $+12.293$ & $0$ & $+12.155$ \\
        $\Delta_{67}=\Delta_{89}$ & $0$ & $+12.011$ & $0$ & $+11.693$ \\
        $\kappa_{11}$ & $0$ & $+3.8762\times 10^{-6}$ & $0$ & $+5.1011\times 10^{-6}$ \\
        $\kappa_{22}$ & $0$ & $+3.8762\times 10^{-6}$ & $0$ & $+5.0102\times 10^{-6}$ \\
        $\kappa_{33}$ & $0$ & $+0.078913$ & $0$ & $+0.098773$ \\
        $\kappa_{12}$ & $0$ & $+0.0017234$ & $0$ & $+0.0020923$ \\
        $\kappa_{13}=\kappa_{23}$ & $0$ & $+0.52751$ & $0$ & $+0.66061$ \\
        $\kappa_{45}$ & $0$ & $-0.0017195$ & $0$ & $-0.0020872$ \\
        $\kappa_{67}=\kappa_{89}$ & $0$ & $-0.48835$ & $0$ & $-0.61184$ \\
        \hline
      \end{tabular}
    }
    \caption[Self-consistent parameters for a CFL phase with finite
    $m_s$]{\label{tab:CFLMs} Parameters required for a self-consistent
      parity even CFL solution in the presence of a strange quark
      mass.  These values correspond to the dispersions shown in
      Figure~\ref{fig:MSDispersions}.  All values are in MeV.  The
      first column labelled ``Bare'' gives the fixed bare parameters
      that enter the Hamiltonian~(\ref{eq:H}).  The column labelled
      ``Correction'' is the contribution from the self-energy.  The
      sum of the columns is the value that enters the quadratic
      Hamiltonian~(\ref{eq:H0}).  For example, the right set of data
      (just slightly before the CFL/gCFL transition) has a bare
      (current) strange quark mass of $80$ MeV.  This corresponds
      to a constituent quark mass of $80+64 \approx 144$ MeV.
      (Note that there is a slight difference for the blue constituent
      quark masses because of the presence of the coloured chemical
      potential $\mu_8$ required to enforce neutrality.)}
  \end{center}
\end{table*}
\begin{table*}[ht]
  \begin{center}
    {\small
      \begin{tabular}{||r||l|l||l|l||}
        \hline
        & \multicolumn{2}{c||}{$M_s^2/(2\mu) = 0.50\mu_Y^c$} 
        & \multicolumn{2}{c||}{$M_s^2/(2\mu) = 0.84\mu_Y^c$}\\
        \cline{2-5}
        Param. & Bare & Correction & Bare & Correction\\
        \hline
        $\mu_B/3$ & $+549.93$ & $-48.951$ & $+549.93$ & $-48.09$ \\
        $\mu_3$ & $-6.6057$ & $-0.00000$ & $-13.002$ & $-8.7\times 10^{-8}$ \\
        $\mu_8$ & $-3.3029$ & $-0.00000$ & $-6.5008$ & $-4.35\times 10^{-8}$ \\
        $\mu_{f}$ & $0$ & $-0.53978$ & $0$ & $-1.0238$ \\
        $\mu_{e3}$ & $0$ & $+0.023555$ & $0$ & $+0.025582$ \\
        $\mu_{e8}$ & $0$ & $+0.0078516$ & $0$ & $+0.0085274$ \\
        $\mu_{f3}$ & $0$ & $+0.035852$ & $0$ & $+0.066184$ \\
        $\mu_{f8}$ & $0$ & $+0.011951$ & $0$ & $+0.022061$ \\
        $\mu_{12}=\mu^5_{18i}$ & $0$ & $+0.014623$ & $0$ & $-0.026924$ \\
        $\mu_{13}=\mu^5_{19i}$ & $0$ & $+0.11615$ & $0$ & $+0.16265$ \\
        $\mu_{23}=\mu_{89}$ & $0$ & $+0.044694$ & $0$ & $+0.042908$ \\
        $\mu^5_{28i}$ & $0$ & $+0.086431$ & $0$ & $+0.11903$ \\
        $\mu^5_{29i}=\mu^5_{38i}$ & $0$ & $+0.044644$ & $0$ & $+0.042927$ \\
        $\mu^5_{39i}$ & $0$ & $-0.001146$ & $0$ & $-0.047235$ \\
        $\mu^5_{46i}$ & $0$ & $+0.11528$ & $0$ & $+0.21622$ \\
        $m_{ug}=-m_{ub}$ & $0$ & $+0.0077476$ & $0$ & $+0.0093125$ \\
        $m_{dr}$ & $0$ & $+0.17093$ & $0$ & $+0.21503$ \\
        $m_{dg}=m_{db}$ & $0$ & $+0.15529$ & $0$ & $+0.19536$ \\
        $m_{sr}$ & $+61.637$ & $+50.279$ & $+85$ & $+69.188$ \\
        $m_{sg}=m_{sb}$ & $+61.637$ & $+50.335$ & $+85$ & $+69.301$ \\
        $m_{12}=-m^5_{18i}$ & $0$ & $-0.010709$ & $0$ & $-0.013162$ \\
        $m_{13}=-m^5_{19i}$ & $0$ & $+0.029548$ & $0$ & $+0.053931$ \\
        $m_{23}=m_{89}$ & $0$ & $+0.0134$ & $0$ & $+0.017695$ \\
        $m^5_{29i}=-m^5_{38i}$ & $0$ & $-0.012525$ & $0$ & $-0.01659$ \\
        $m^5_{46i}$ & $0$ & $-0.0077707$ & $0$ & $-0.009337$ \\
        \hline
        $\Delta_{11}$ & $0$ & $+0.60701$ & $0$ & $+0.53896$ \\
        $\Delta_{22}=-\Delta_{88}$ & $0$ & $+0.30409$ & $0$ & $+0.28263$ \\
        $\Delta_{33}=-\Delta_{99}$ & $0$ & $+0.32668$ & $0$ & $+0.32494$ \\
        $\Delta_{12}=-\Delta^5_{18i}$ & $0$ & $+9.4718$ & $0$ & $+9.6392$ \\
        $\Delta_{13}=-\Delta^5_{19i}$ & $0$ & $+8.614$ & $0$ & $+8.0104$ \\
        $\Delta_{23}=-\Delta_{89}$ & $0$ & $+12.282$ & $0$ & $+11.726$ \\
        $\Delta_{45}=-\Delta^5_{56i}$ & $0$ & $-9.0411$ & $0$ & $-9.2445$ \\
        $\Delta_{67}=-\Delta^5_{47i}$ & $0$ & $-8.1684$ & $0$ & $-7.5847$ \\
        $\Delta^5_{28i}$ & $0$ & $-0.30382$ & $0$ & $-0.28208$ \\
        $\Delta^5_{29i}=\Delta^5_{38i}$ & $0$ & $-0.31799$ & $0$ & $-0.31604$ \\
        $\Delta^5_{39i}$ & $0$ & $-0.32038$ & $0$ & $-0.31321$ \\
        $\kappa_{22}=-\kappa_{88}$ & $0$ & $-1.7029\times 10^{-5}$ & $0$ & $-1.3167\times 10^{-5}$ \\
        $\kappa_{33}=-\kappa_{99}$ & $0$ & $-0.040639$ & $0$ & $-0.055178$ \\
        $\kappa_{12}=\kappa^5_{18i}$ & $0$ & $-0.00066095$ & $0$ & $-0.00090186$ \\
        $\kappa_{13}=\kappa^5_{19i}$ & $0$ & $-0.36095$ & $0$ & $-0.46324$ \\
        $\kappa_{23}=-\kappa_{89}$ & $0$ & $-0.50626$ & $0$ & $-0.66374$ \\
        $\kappa_{45}=-\kappa^5_{56i}$ & $0$ & $+0.00075037$ & $0$ & $+0.001152$ \\
        $\kappa_{67}=\kappa^5_{47i}$ & $0$ & $+0.33313$ & $0$ & $+0.4268$ \\
        $\kappa^5_{29i}=-\kappa^5_{38i}$ & $0$ & $-0.019786$ & $0$ & $-0.026852$ \\
        \hline
      \end{tabular}
      \caption[Self-consistent parameters for a CFLK$0$ phase with finite
      $m_s$]{\label{tab:CFLK0Ms} Parameters required for a
        self-consistent CFLK$0$ solution in the presence of a strange
        quark mass.  These values correspond to the dispersions shown in
        Figure~\ref{fig:MSK0Dispersions}.  All values are in MeV.}
    }
  \end{center}
\end{table*}
\end{document}